\documentclass{article}

\usepackage{arxiv}

\usepackage[utf8]{inputenc} 
\usepackage[T1]{fontenc}    
\usepackage{hyperref}       
\usepackage{url}            
\usepackage{booktabs}       
\usepackage{amsfonts}       
\usepackage{nicefrac}       
\usepackage{microtype}      
\usepackage{lipsum}		
\usepackage{graphicx}
\usepackage{natbib}
\usepackage{doi}

\usepackage{preamble}
\usepackage{enumitem}

\title{Optimal multi-wave validation of secondary use data with outcome and exposure misclassification}

\author{
\href{https://orcid.org/0000-0001-5380-2427}{\includegraphics[scale=0.06]{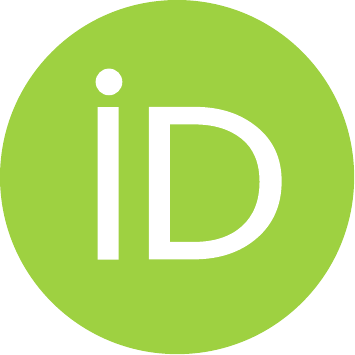}\hspace{1mm}Sarah C.~Lotspeich} \\
	Department of Biostatistics\\
	Vanderbilt University Medical Center\\
	Nashville, TN, U.S.A.\\
	Department of Biostatistics\\
	University of North Carolina at Chapel Hill\\
	Chapel Hill, NC, U.S.A.\\
	\texttt{lotspes@wfu.edu} \\
	\And
	Gustavo G. C.~Amorim \\
	Department of Biostatistics\\
	Vanderbilt University Medical Center\\
	Nashville, TN, U.S.A.\\
	\And 
	Pamela A.~Shaw \\
	Department of Biostatistics, Epidemiology, and Informatics \\
	University of Pennsylvania \\
	Philadelphia, PA, U.S.A. \\
	Biostatistics Unit \\
	Kaiser Permanente Washington Health Research Institute \\
	Seattle, WA, U.S.A.
	\And 
	Ran~Tao\thanks{These authors jointly supervised this work.} \\
	Department of Biostatistics\\
	Vanderbilt University Medical Center\\
	Nashville, TN, U.S.A.\\
	Vanderbilt Genetics Institute \\
	Vanderbilt University Medical Center\\
	Nashville, TN, U.S.A.\\
	\And 
	Bryan E.~Shepherd$^*$ \\
	Department of Biostatistics\\
	Vanderbilt University Medical Center\\
	Nashville, TN, U.S.A.
}

\renewcommand{\shorttitle}{Optimal multi-wave validation of secondary use data}

\hypersetup{
pdftitle={Optimal multi-wave validation of secondary use data},
pdfsubject={q-bio.NC, q-bio.QM},
pdfauthor={Sarah C.~Lotspeich, Gustavo G. C.~Amorim, Pamela A.~Shaw, Ran~Tao, Bryan E.~Shepherd},
pdfkeywords={Data audits, HIV/AIDS, Likelihood estimation, Measurement error, Two-phase designs},
}

\begin{document}
\maketitle

\begin{abstract}
	The growing availability of observational databases like electronic health records (EHR) provides unprecedented opportunities for secondary use of such data in biomedical research. However, these data can be error-prone and need to be validated before use. It is usually unrealistic to validate the whole database due to resource constraints. A cost-effective alternative is to implement a two-phase design that validates a subset of patient records that are enriched for information about the research question of interest. Herein, we consider odds ratio estimation under differential outcome and exposure misclassification. We propose optimal designs that minimize the variance of the maximum likelihood odds ratio estimator. We develop a novel adaptive grid search algorithm that can locate the optimal design in a computationally feasible and numerically accurate manner. Because the optimal design requires specification of unknown parameters at the outset and thus is unattainable without prior information, we introduce a multi-wave sampling strategy to approximate it in practice. We demonstrate the efficiency gains of the proposed designs over existing ones through extensive simulations and two large observational studies. We provide an R package and Shiny app to facilitate the use of the optimal designs.
\end{abstract}


\section{INTRODUCTION}

The ever-growing trove of patient information in observational databases, like electronic health records (EHR), provides unprecedented opportunities for biomedical researchers to investigate associations of scientific and clinical interest. However, these data are usually error-prone since they are ``secondary use data,'' i.e., they were not primarily created for research purposes (Safran et al., 2007). Ignoring the errors can yield biased results (Giganti et al., 2019), and the interpretation, dissemination, or implementation of such results can be detrimental to the very patients whom the analysis sought to help.

To assess the quality of secondary use data, validation studies have been carried out, wherein trained auditors compare clinical source documents (e.g., paper medical records) to database values and note any discrepancies between them (Duda et al., 2012). The Vanderbilt Comprehensive Care Clinic (VCCC) is an outpatient facility in Nashville, Tennessee that provides care for people living with HIV/AIDS (PLWH). Since investigators at the VCCC extract EHR data for research purposes, the VCCC validates all key study variables for all patients in the EHR. The VCCC data have demonstrated the importance of data validation, as estimates using the fully validated data often differ substantially from those using the original, unvalidated data extracted from the EHR (Giganti et al., 2020).

However, validating entire databases can be cost-prohibitive and unattainable; in the VCCC, full-database validation of around 4000 patients costs over US\$60,000 annually. A cost-effective alternative is a two-phase design (White, 1982), or partial data audit, wherein one collects the original, error-prone data in Phase I and then uses this information to select a subset of records for validation in Phase II. This design greatly reduces the cost associated with data validation and has been implemented in cohorts using routinely collected data, like the Caribbean, Central, and South America network for HIV Epidemiology (CCASAnet) (McGowan et al., 2007).

CCASAnet is a large (${\sim}\num{50000}$ patients), multi-national HIV clinical research collaboration. Clinical sites in CCASAnet routinely collect important variables, and these site-level data are subsequently compiled into a collaborative CCASAnet database that is used for research. One interesting question for CCASAnet investigators is whether patients treated for tuberculosis (TB) are more likely to have better treatment outcomes if their TB diagnosis was bacteriologically confirmed. TB is difficult to diagnose and treat among PLWH, and some studies suggest that those treated for TB without a definitive diagnosis are more likely to subsequently die (Crabtree-Ramirez et al., 2019). Key study variables can be obtained from the CCASAnet database, but the outcome and exposure, successful treatment completion and bacteriological confirmation, respectively, can be misclassified in the database. For more than a decade, partial data audits have been performed to ensure the integrity of the CCASAnet research database (Duda et al., 2012; Giganti et al., 2019; Lotspeich et al., 2020), and plans are currently underway to validate these TB study variables on a subset of records in the near future. Site-stratified random sampling has been the most common selection mechanism thus far, including a 2009--2010 audit of the TB variables. Now, we are interested in developing optimal designs that select subjects who are most informative for our study of the association between bacteriologic confirmation and treatment completion to answer this question with the best possible precision.

Statistical methods have been proposed to analyze data from two-phase studies like this, correcting for binary outcome misclassification and exposure errors {\color{black}simultaneously}. These methods can largely be grouped into likelihood- or design-based estimators. The former include the maximum likelihood estimator (MLE) (Tang et al., 2015) and semiparametric maximum likelihood estimator (SMLE) (Lotspeich et al., 2021), while the latter include the inverse probability weighted (IPW) estimator (Horvitz \& Thompson, 1952) and generalized raking/augmented IPW estimator (Deville, Sarndal \& Sautory, 1993; Robins, Rotnitzky \& Zhao, 1994; Oh et al., 2021b). Likelihood-based estimators are fully efficient when {\color{black} all models (i.e., analysis and misclassification models) are} correctly specified, while design-based estimators tend to be more robust since they make fewer distributional assumptions {\color{black}(i.e., do not require specification of a misclassification model)}. We focus on full-likelihood estimators because we have full-cohort information and they offer greater efficiency. Theoretical properties and empirical comparisons of these estimators have been discussed in detail before (e.g., McIsaac \& Cook, 2014; Lotspeich et al., 2021). Thus, in this paper, we focus on designs, rather than estimation, for two-phase studies.

Given the resource constraints imposed upon data audits, efficient designs that target the most informative patients are salient. Closed-form solutions exist for the optimal sampling proportions {\color{black}to minimize the variances} for some design-based estimators under settings with {\color{black}outcome (e.g., Pepe, Reilly \& Fleming, 1994)} or exposure error alone ({\color{black} e.g.,} Reilly \& Pepe, 1995; McIsaac \& Cook, 2014; Chen \& Lumley, 2020). Optimal designs for likelihood-based estimators have also been considered in the setting of exposure errors {\color{black}alone, although the variances of these estimators do not lend themselves to closed-form solutions} unless additional assumptions are made (Breslow \& Cain, 1988; Holcroft \& Spiegelman, 1999; McIsaac \& Cook, 2014; Tao, Zeng \& Lin, 2020).

{\color{black}Still, }optimal designs have yet to be developed for two-phase studies with a misclassified binary outcome \textit{and} exposure error, as needed for the CCASAnet TB study. Existing designs for our setting are limited to case-control (CC*) or balanced case-control (BCC*) designs based on the unvalidated outcome and exposure (Breslow \& Cain, 1988) (we use the * here to differentiate these designs, which are based on error-prone data, from their traditional counterparts). {\color{black}In fact, 
many of the designs proposed for expensive variables (a setting similar to measurement error) are just CC* or BCC* sampling, or variations of them, since they target a particular prevalence for the variable of interest (Tan \& Heagerty, 2022) or ``balanced'' numbers from each of the Phase I strata to sample in Phase II (White, 1982; Wang et al., 2017), respectively. While these designs are practical and can offer efficiency gains over simple random sampling, they were not derived to be optimal for any specific target parameter.} Our goal is to compute optimal designs for likelihood-based estimators in the unaddressed setting of binary outcome and exposure misclassification.

Regardless of the estimator, optimal designs share common challenges; in particular, they require specification of unknown parameters. To overcome this, multi-wave strategies have been proposed that estimate the unknown parameters with an internal pilot study and use this information to approximate the optimal design (McIsaac \& Cook, 2015; Chen \& Lumley, 2020; Han et al., 2020; {\color{black}Shepherd et al., 2022}). Instead of selecting one Phase II subsample, multi-wave designs allow iterative selection in two or more waves of Phase II. This way, each wave gains information from those that came before it. So far, multi-wave designs have only been used to adapt optimal designs for design-based estimators. We focus on designing multi-wave validation studies to improve the statistical efficiency of likelihood-based estimators {\color{black} under outcome and exposure misclassification}.

Based on the asymptotic properties of the two-phase MLE for logistic regression, we derive the optimal validation study design to minimize the variance of the log odds ratio (OR) under differential outcome and exposure misclassification. In the absence of a closed-form solution, we devise an adaptive grid search algorithm that can locate the optimal design in a computationally feasible and numerically accurate manner. Because the optimal design requires specification of unknown parameters at the outset and thus is unattainable without prior information, we introduce a multi-wave sampling strategy to approximate it in practice. Through extensive simulations, the proposed optimal designs are compared to CC* and BCC* sampling. Notable gains in efficiency can be seen not only with the optimal design, but also with the multi-wave approximation to it. Using the VCCC data, we evaluate the efficiency of various designs validating different subsets of the EHR data and compare them to the fully validated, full-cohort analysis. Finally, we implement our approach to design the next round of CCASAnet audits.

\section{METHODS}

\subsection{Model and data}\label{Paper3_Methods_Model&Data}
Consider a binary outcome, $Y$, binary exposure, $X$, and covariates $\pmb{Z}$ that are assumed to be related through the logistic regression model $P(Y=1|X,\pmb{Z}) = [1 + \exp\{-(\beta_0 + \beta X + \pmb{Z}\pmb{\beta}_z)\}]^{-1}$. Instead of $Y$ and $X$, error-prone measures $Y^*$ and $X^*$, respectively, are available in an observational database. Covariates $\pmb{Z}$ are also available and error-free. An audit sample of size $n$ of the $N$ subjects in the database ($n < N$) will have their data validated. Let $V_i=1$ if subject $i$ ($i = 1, \dots, N$) is audited and $0$ otherwise. The joint distribution of a complete observation is $P(V,Y^*,X^*,Y,X,\pmb{Z})=$
\begin{align}
& P(V|Y^*,X^*,\pmb{Z})P(Y^*|X^*,Y,X,\pmb{Z})P(X^*|Y,X,\pmb{Z})P(Y|X,\pmb{Z})P(X|\pmb{Z})P(\pmb{Z}), \label{joint}
\end{align}
where $P(V|Y^*,X^*,\pmb{Z})$ is the validation sampling probability; $P(Y|X,\pmb{Z})$ is the logistic regression model of primary interest; $P(Y^*|X^*,Y,X,\pmb{Z})$ and $P(X^*|Y,X,\pmb{Z})$ are the outcome and exposure misclassification mechanisms, respectively; $P(X|\pmb{Z})$ is the conditional probability of $X$ given $\pmb{Z}$; and $P(\pmb{Z})$ is the marginal density of $\pmb{Z}$. Sampling (i.e., $V$) is assumed to depend only on Phase I variables ($Y^*$, $X^*$, $\pmb{Z}$), so $(Y,X)$ are missing at random (MAR) for unaudited subjects.

Equation~\eqref{joint} captures the most {\color{black}general situation with} complex differential misclassification in the outcome and exposure {\color{black} and without any assumptions of independence between variables}, but it addresses other common settings as special cases. For classical scenarios of outcome or exposure misclassification alone, set $X^* = X$ or $Y^* = Y$, respectively. For nondifferential misclassification, let $P(Y^*|X^*, Y, X ,\pmb{Z}) = P(Y^*|Y, \pmb{Z})$ or $P(X^*|Y,X,\pmb{Z}) = P(X^*|X,\pmb{Z})$ (Keogh et al., 2020). {\color{black}If one has more specific knowledge perhaps from a previous audit, scientific context, or the literature, the models can be further customized. For example, if the exposure and covariates are assumed to be independent, then $P(X|\pmb{Z}) = P(X)$. Importantly, these customizations do not affect the derivations of the optimal design that follow.}

All observations ($V_i,Y^*_i,X^*_i,Y_i,X_i,\pmb{Z}_i$) ($i = 1, \dots, N$) are assumed to be i.i.d. following Equation~\eqref{joint}. The necessary unknowns in Equation~\eqref{joint} -- specifically, $P(Y^*_i|X^*_i,Y_i,X_i,\pmb{Z}_i)$, $P(X^*_i|Y_i,X_i,\pmb{Z}_i)$, and $P(X_i|\pmb{Z}_i)$ -- are assumed to follow additional logistic regression models. Model parameters are denoted together by $\pmb{\theta}$; since we focus on estimating $\beta$, all other nuisance parameters are denoted by $\pmb{\eta}$ such that $\pmb{\theta} = \left(\beta, \pmb{\eta}^{\rm T}\right)^{\rm T}$. Given that $(Y_i,X_i)$ are incompletely observed, the observed-data log-likelihood for $\pmb{\theta}$ is $l_{N}(\pmb{\theta}) = $
\begin{align}
&\sum_{i=1}^{N}V_i\left\{\log P(Y_i^*|X_i^*,Y_i,X_i,\pmb{Z}_i) + \log P(X_i^*|Y_i,X_i,\pmb{Z}_i) + \log P(Y_i|X_i,\pmb{Z}_i) + \log P(X_i|\pmb{Z}_i) \right\} \nonumber \\
& + \sum_{i=1}^{N}(1-V_i)\log\left\{\sum_{y=0}^{1}\sum_{x=0}^{1}P(Y_i^*|X_i^*,y,x,\pmb{Z}_i)P(X_i^*|y,x,\pmb{Z}_i)P(y|x,\pmb{Z}_i)P(x|\pmb{Z}_i)\right\}. \label{od_ll}
\end{align}
The distribution of $V$ can be omitted because the Phase II variables are MAR. {\color{black} In other words, because $P(V|Y^*,X^*,\pmb{Z})$ is fully observed (in fact, fixed by design) it would only rescale $l_{N}(\pmb{\theta})$ by a constant, so omitting it from Equation~\eqref{od_ll} does not affect parameter estimation.} The MLE $\widehat{\pmb{\theta}} = (\hat{\beta}, \hat{\pmb{\eta}}^{\rm T})^{\rm T}$ can be obtained by maximizing Equation \eqref{od_ll} (Tang et al., 2015). Our optimal design will obtain the most efficient MLE for $\beta$, the conditional log OR for $X$ on $Y$. 

\subsection{Optimal design}\label{Paper3_Methods_Design}

Under standard asymptotic theory with $N \rightarrow \infty$ and $n/N \rightarrow P(V=1) >0$, $\sqrt{N}(\widehat{\pmb{\theta}} - \pmb{\theta}) \rightsquigarrow \pmb{N}(\pmb{0},\mathcal{I}(\pmb{\theta})^{-1})$
where $\rightsquigarrow$ represents convergence in distribution and $\pmb{N}(\pmb{0},\mathcal{I}(\pmb{\theta})^{-1})$ is a multivariate normal distribution with mean $\pmb{0}$ and variance equal to the inverse of the Fisher information, $\mathcal{I}(\pmb{\theta})$. Partition $\mathcal{I}(\pmb{\theta})$ as
\begin{align}
\mathcal{I}(\pmb{\theta}) 
= \begin{bmatrix}
\mathcal{I}(\beta,\beta) & \mathcal{I}(\beta,\pmb{\eta})^{\rm T} \\
\mathcal{I}(\beta,\pmb{\eta}) & \mathcal{I}(\pmb{\eta,\eta}) \\
\end{bmatrix}. \nonumber 
\end{align}
The optimal design aims to minimize ${\rm Var}(\hat{\beta})$, which can be expressed as
\begin{align}
{\rm Var}(\hat{\beta}) &= N^{-1}\left\{\mathcal{I}(\beta,\beta) - \mathcal{I}(\beta,\pmb{\eta})^{\rm T}\mathcal{I}(\pmb{\eta},\pmb{\eta})^{-1}\mathcal{I}(\beta,\pmb{\eta}) \right\}^{-1} \label{V_beta}.
\end{align}

The elements of $\mathcal{I}(\pmb{\theta})$ are expectations taken {\color{black}with respect to the complete data, following from the joint density in Equation~\eqref{joint}. Thus, they can be expressed} as functions of the sampling probabilities, $\pi_{y^*x^*\pmb{z}} \equiv P(V=1|Y^*=y^*,X^*=x^*,\pmb{Z}=\pmb{z})$, and model parameters, $\pmb{\theta}$. That is, for elements $\theta_j$ and $\theta_{j'}$  of $\pmb{\theta}$, $\mathcal{I}(\theta_j,\theta_{j'}) =$
\begin{align}
& \sum_{y^*=0}^{1}\sum_{x^*=0}^{1}\sum_{\pmb{z}}\pi_{y^*x^*\pmb{z}}\sum_{y=0}^{1}\sum_{x=0}^{1}S^{v}(\theta_j;y^*,x^*,y,x,\pmb{z})S^{v}(\theta_{j'};y^*,x^*,y,x,\pmb{z})P(y^*,x^*,y,x,\pmb{z}) \nonumber \\
& + \sum_{y^*=0}^{1}\sum_{x^*=0}^{1}\sum_{\pmb{z}}(1 - \pi_{y^*x^*\pmb{z}})S^{\bar{v}}(\theta_j;y^*,x^*,\pmb{z})S^{\bar{v}}(\theta_{j'};y^*,x^*,\pmb{z})\sum_{y=0}^{1}\sum_{x=0}^{1}P(y^*,x^*,y,x,\pmb{z}), \label{I_theta_theta}
\end{align}
where $S^{v}(\cdot)$ and $S^{\bar{v}}(\cdot)$ are the score functions of validated and unvalidated subjects, respectively, {\color{black}and $P(Y^*,X^*,Y,X,\pmb{Z})$ is the joint density of the error-prone and error-free variables (see Appendix for details)}. The sampling strata are {\color{black}defined by $Y^*$, $X^*$, and 
$\pmb{Z}$.} 

{\color{black} Note that Equation~\eqref{I_theta_theta} implicitly assumes that the covariates $\bZ$ are discrete. As will be seen in our applications, this is sometimes the case (e.g., $Z$ is study site) but certainly not always (e.g., $Z$ is a continuous lab value). If the covariates 
are continuous or multi-dimensional with many categories, one will need to simplify them 
to create sampling strata. Specifically, the covariates 
may need to be discretized or have their dimensions reduced to make sampling feasible; such a strategy has been employed by others (e.g., Lawless, Kalbfleisch \& Wild, 1999; Zhou et al., 2002; Tao, Zeng \& Lin, 2020; Han et al., 2020). The resulting optimal design based on the simplified covariates may no longer be optimal for minimizing the variance based on the full, unsimplified covariates, 
but in most scenarios it should be a good approximation to the optimal design and better than classical designs like BCC* (Tao, Zeng \& Lin, 2020). Moreover, the resulting discretized design should converge to the optimal one as the number of strata increases. In practice, there also exists a trade-off between the pursuit of optimality (with more strata) and ease of implementation (with fewer strata).}

We see from Equations~\eqref{V_beta} and \eqref{I_theta_theta} that the optimal design corresponds to $\{\pi_{y^*x^*\pmb{z}}\}$ that minimizes ${\rm Var}(\hat{\beta})$ under the constraint
\begin{align}
\sum_{y^*=0}^{1}\sum_{x^*=0}^{1}\sum_{\pmb{z}}\pi_{y^*x^*\pmb{z}}N_{y^*x^*\pmb{z}}\equiv\sum_{y^*=0}^{1}\sum_{x^*=0}^{1}\sum_{\pmb{z}}n_{y^*x^*\pmb{z}}=n,  \label{constraint}
\end{align}
where $N_{y^*x^*\pmb{z}}$ and $n_{y^*x^*\pmb{z}}$ are the sizes of the stratum ($Y^*_i=y^*$, $X^*_i=x^*$, $\pmb{Z}_i = \pmb{z}$) in Phase I and Phase II, respectively. Because $\{N_{y^*x^*\pmb{z}}\}$ are fixed, finding the optimal values of $\{\pi_{y^*x^*\pmb{z}}\}$ is equivalent to finding the optimal values of  $\{n_{y^*x^*\pmb{z}}\}$. Unfortunately, this constrained optimization problem does not have a closed-form solution, so we devise an adaptive grid search algorithm to find the optimal values of $\{n_{y^*x^*\pmb{z}}\}$.

\subsection{Adaptive grid search algorithm}\label{Paper3_Methods_GridSearch}

The challenge at hand is one of combinatorics: of all the candidate designs that satisfy the audit size constraint and are supported by the available Phase I data (i.e., the stratum sizes $\{N_{y^*x^*\pmb{z}}\}$), {\color{black}we need to find the one that minimizes ${\rm Var}(\hat{\beta})$. To locate the optimal design, we develop an adaptive grid search algorithm. Specifically, a series of grids are constructed at iteratively finer scales and over more focused candidate design spaces to locate the optimal design.} The adaptive nature of our algorithm is necessitated by the {\color{black}dimensional} explosion of the grid as the Phase I and Phase II sample sizes and number of sampling strata increase.

Let $K$ denote the number of sampling strata and let $m$ denote the minimum number that must be sampled in a stratum; $m$ is needed to avoid degenerate optimal designs {\color{black} where one or more sampling strata are empty, rendering the estimation of the conditional distribution of Phase I given Phase II data impossible} (Breslow \& Cain, 1988). 
At the first iteration of the grid search, we form the grid $\pmb{G}^{(1)}$ with candidate designs comprised of stratum sizes {\color{black}$n_{y^*x^*\pmb{z}}^{(1)}$} that satisfy constraints \eqref{constraint} and 
\begin{align}
\min(m,N_{y^*x^*\pmb{z}}) \leq n_{y^*x^*\pmb{z}}{\color{black}^{(1)}} \leq \min(n-Km+m, N_{y^*x^*\pmb{z}}),
\label{cons1st}
\end{align}
i.e., candidate stratum sizes fall between the minimum, $m$, and the maximum after minimally allocating to all $K$ strata, $n - Km + m$ (or the full stratum size $N_{y^*x^*\pmb{z}}$ if either of these is not possible). The number of subjects $n_{y^*x^*z}$ in each stratum varies by a fixed quantity $s^{(1)}$ between candidate designs. For example, we might consider sampling {\color{black}$n_{y^*x^*\pmb{z}}^{(1)} \in \{10, 25, \dots, 100\}$ subjects if $s^{(1)} = 15$}, $m=10$, and $N_{y^*x^*\pmb{z}} = 100$. We then calculate ${\rm Var}(\hat{\beta})$ under each candidate design to identify the best one, i.e., the one that yields the smallest ${\rm Var}(\hat{\beta})$. Given the large space of legitimate designs in this initial search, we want to choose a reasonably large $s^{(1)}$ to keep the dimension of $\pmb{G}^{(1)}$ manageable. Clearly, starting with a large $s^{(1)}$ will lead to a rough estimate of the optimal design, but this will be refined in subsequent iterations.

At the $t$th iteration ($t > 1$), we form the grid $\pmb{G}^{(t)}$ around the ``$s^{(t-1)}$-neighborhood" of the best design from the $(t-1)$th iteration whose best stratum sizes are denoted by $\{n^{(t-1)}_{y^*x^*\pmb{z}}\}$. That is, we construct $\pmb{G}^{(t)}$ from candidate designs that satisfy constraints \eqref{constraint} and
\begin{align}
\max(n^{(t-1)}_{y^*x^*\pmb{z}}-s^{(t-1)},m) \leq n_{y^*x^*\pmb{z}}{\color{black}^{(t)}} \leq \min(n^{(t-1)}_{y^*x^*z}+s^{(t-1)}, N_{y^*x^*\pmb{z}}). \label{constth} 
\end{align}
This constraint is a refined version of \eqref{cons1st}, since we focus on a narrower space of designs surrounding the previous iteration's best design. Once again, the stratum sizes $\{n_{y^*x^*\pmb{z}}^{(t)}\}$ vary by multiples of $s^{(t)}$ between candidate designs. We adaptively choose $s^{(t)} < s^{(t-1)}$ such that the grids $\{\pmb{G}^{(t)}\}$ become finer and finer during the iterative process. {\color{black}The choice of the step sizes $s^{(1)}, \dots, s^{(T)}$ will determine the computation time to complete the algorithm, but the grid search appears robust to these choices (Web Appendix A).} We stop at $s^{(T)}=1$, meaning that the final search was conducted at the 1-person level, and the best design at the last iteration ($T$) is the optimal design, which we call the optMLE. 

\begin{figure}
\begin{subfigure}[b]{\textwidth}
\includegraphics[width=\linewidth]{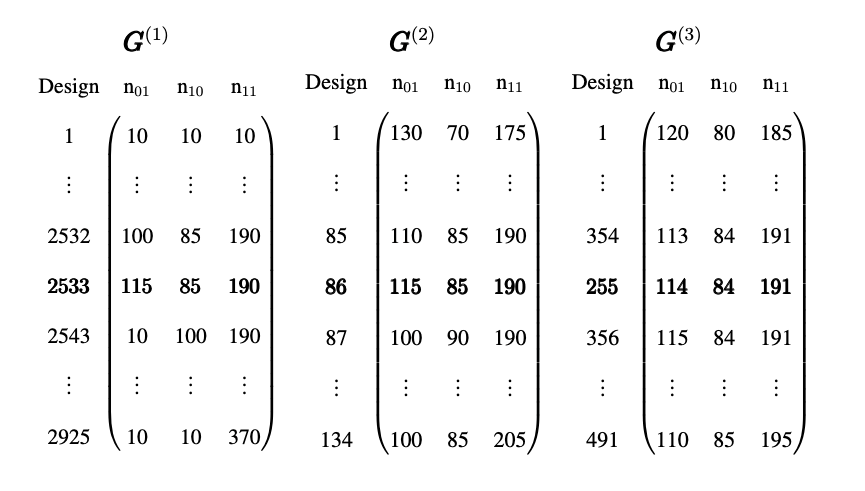}
\caption{Matrix representation}
\end{subfigure}
\begin{subfigure}[b]{\textwidth}
\includegraphics[width=\linewidth]{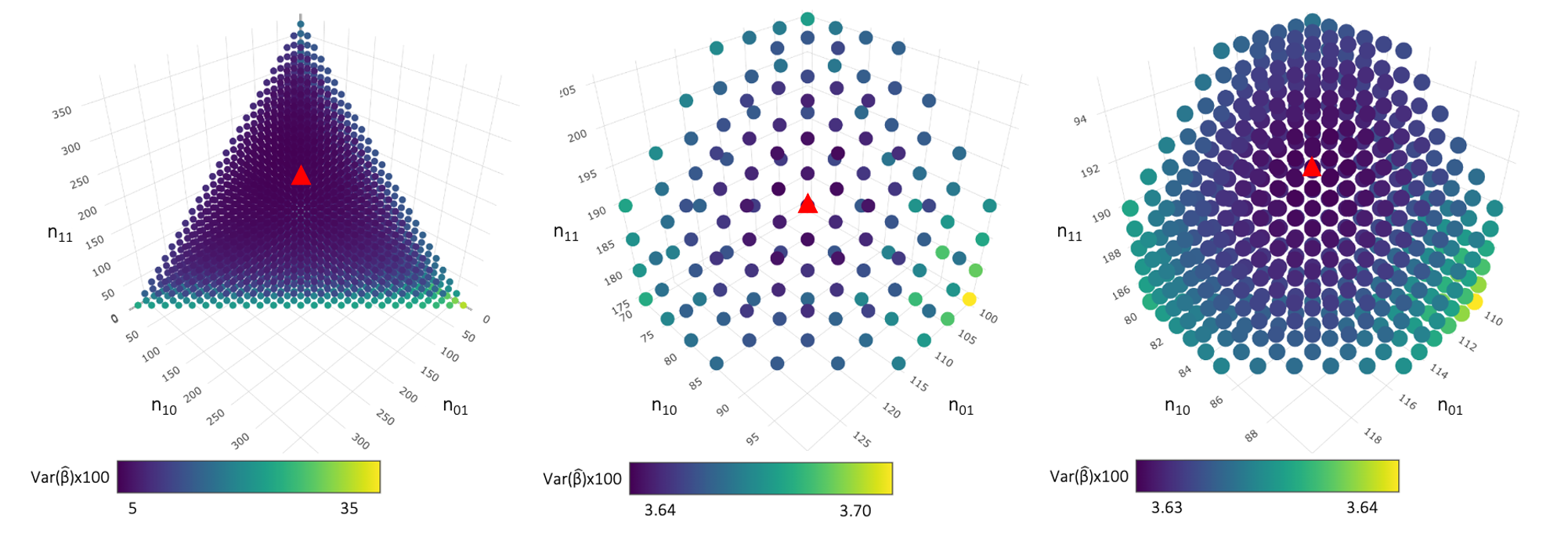}
\caption{Graphical representation}
\end{subfigure}
\caption{Matrix and graphical representations of a three-iteration adaptive grid search for a validation study of $n = 400$. In (a), the bold row indicates the design achieving the lowest ${\rm Var}(\hat{\beta)}$; in (b) the triangle does. $n_{00}$ can be omitted because it is determined by constraint (5).}
\label{fig:ex_grid}
\end{figure}

Figure~\ref{fig:ex_grid} depicts a schematic diagram of an adaptive grid search with $T = 3$ iterations. In this hypothetical example, the Phase I sample size is $N = \num{10000}$ and there are $K = 4$ strata defined by ($Y^*$, $X^*$) with Phase I stratum sizes $\{N_{y^*x^*}\} = (N_{00}, N_{01}, N_{10}, N_{11})=$ (5297, 1130, 2655, 918). The aim is to select $n = 400$ subjects for data validation in Phase II. Based on our simulations (discussed in Section~\ref{Paper3_Res_Both}), we set $m=10$. Assume that reliable parameter estimates are available from a previous data audit which can be used in the grid search. At the first iteration, we construct $\pmb{G}^{(1)}$ with candidate designs that satisfy constraints \eqref{constraint} and \eqref{cons1st}, varying the stratum sizes $\{n_{y^*x^*}\}$ by multiples of $s^{(1)} = 15$ between designs. ${\rm Var}(\hat{\beta})$ is minimized at $3.6283\times 10^{-4}$ under the candidate design with Phase II stratum sizes $\{n^{(1)}_{y^*x^*}\} \equiv (n_{00}^{(1)}, n_{01}^{(1)}, n_{10}^{(1)}, n_{11}^{(1)}) = (10, 115, 85, 190)$. At the second iteration, we form the grid $\pmb{G}^{(2)}$ from the 15-person-neighborhood around $\{n^{(1)}_{y^*x^*}\}$ such that the candidate designs satisfy constraints \eqref{constraint} and \eqref{constth}, with stratum sizes varied by multiples of $s^{(2)}=5$ between designs. ${\rm Var}(\hat{\beta})$ is minimized under the same design as in the first iteration, i.e., $\{n^{(2)}_{y^*x^*}\}$ = $\{n^{(1)}_{y^*x^*}\}$. At the third and last iteration, we form grid $\pmb{G}^{(3)}$ in the 5-person-neighborhood around $\{n^{(2)}_{y^*x^*}\}$ such that the candidate designs satisfy constraints \eqref{constraint} and \eqref{constth} with stratum sizes varied by multiples of $s^{(3)} = 1$ between designs. ${\rm Var}(\hat{\beta})$ is minimized at $3.6281\times 10^{-4}$ by Phase II sample sizes $\{n^{(3)}_{y^*x^*}\} =$ (11, 114, 84, 191), which is the optMLE design. We note that the minimum variance barely changed between iterations in this toy example; the algorithm proceeds anyway because the stopping rule is defined as the most granular grid search (i.e., a 1-person scale with $s^{(T)} = 1$). In practice, one may use other sensible rules that permit early stops to make the algorithm more computationally efficient, e.g., stop when the minimum variance from successive iterations changes by less than $1\%$. 

\subsection{Multi-wave approximate optimal design}\label{Paper3_Methods_MultiWave}

The optMLE design derived in Section~\ref{Paper3_Methods_Design} relies on model parameters $\pmb{\theta}$, which are unknown at the study outset. If available, historical data from a previous audit could be used to estimate $\pmb{\theta}$. Otherwise, it would be difficult to implement the optMLE design in practice, so we propose a multi-wave sampling strategy to approximate it. Whereas traditional two-phase studies require all design-relevant information to be available in Phase I, multi-wave designs allow sampling to adapt as such information accumulates through multiple sampling waves in Phase II.

{\color{black}In one of the earliest multi-wave papers, McIsaac \& Cook (2015) considered the most and least extreme multi-wave sampling strategies: (i) fully adaptive and (ii) two waves, respectively. Strategy (i) began with a small initial sample, and then the sampling strategy was re-computed after data were collected for each individual selected into Phase II (leading to nearly $n$ waves), while strategy (ii) used just two waves and re-designed the study just once (after the initial sample). McIsaac \& Cook (2015) found that the fully adaptive designs offered great efficiency, but their implementation could be unrealistic in practice; meanwhile, the more practical two-wave strategy offered near-optimal efficiency. Therefore, in this manuscript we primarily consider two waves of sampling, labeled as Phase II(a) and Phase II(b), with the corresponding sample sizes denoted by $n^{(a)}$ and $n^{(b)}$, respectively ($n^{(a)} + n^{(b)}=n$).}

McIsaac \& Cook (2015) also examined different $n^{(a)}:n^{(b)}$ ratios and found that the 1:1 ratio appeared to strike a good balance between (i) the more accurate Phase II(a) estimation of $\pmb{\theta}$ with larger $n^{(a)}$ and (ii) the more flexible design optimization in Phase II(b) with larger $n^{(b)}$. Based on this result, we select $n^{(a)} = n^{(b)} = n / 2$ subjects in each wave. In Phase II(a), we typically select subjects through BCC* sampling; other existing designs could be used depending on the available information (e.g., Oh et al., 2021a). Then, we use the Phase I and Phase II(a) data to compute the preliminary MLE of the parameters, denoted $\widehat{\pmb{\theta}}^{(a)}$, where the validation indicator $V_i = 1$ if subject $i$ ($i = 1, \dots, N$) was sampled in Phase II(a) and $0$ otherwise. Then, we use the grid search algorithm in Section \ref{Paper3_Methods_GridSearch} with $\widehat{\pmb{\theta}}^{(a)}$ to determine the optimal allocation of the remaining subjects in Phase II(b). We call our two-wave approximate optimal design the optMLE-2. Following completion of both waves of validation, the final MLE $\widehat{\pmb{\theta}}$ are obtained by combining data from Phases I, II(a), and II(b) and with redefined validation indicator $V_i = 1$ if subject $i$ ($i = 1, \dots, N$) was audited in either wave of the optMLE-2 design. Thus, ensuing inference is based on $n$ audited and $(N - n)$ unaudited subjects, as with a single wave of audits. 

\section{SIMULATIONS}\label{Paper3_Res}

Our objective with these simulation studies is three-fold: (i) to describe the construction of the optimal designs, since there is not a closed-form; (ii) to demonstrate the efficiency gains of the optimal designs over existing designs; and (iii) to investigate the robustness of the proposed designs to model misspecification. This was explored through settings with varied rates of misclassification (Section~\ref{Paper3_Res_Both}), additional error-free information to incorporate (Section~\ref{Paper3_Res_Both_inclZ}), under model misspecification of the misclassification mechanisms at the design stage (Section~\ref{Paper3_Res_Misspec}), and in special cases with either outcome or exposure misclassification alone (Section~\ref{Paper3_Res_Classical}). 

\subsection{Validation study designs}\label{Paper3_Des}
We compare the performance of five two-phase validation study designs under differential outcome and exposure misclassification. \textit{Simple random sampling (SRS)}: All subjects in Phase I have equal probability of inclusion in Phase II. \textit{CC*}: Subjects are stratified on $Y^*$, and separate random samples of equal size are drawn from each stratum. \textit{BCC*}: Subjects are jointly stratified on ($Y^*, X^*$) or ($Y^*, X^*$, $Z$), and separate random samples of equal size are drawn from each stratum. \textit{optMLE}: Subjects are jointly stratified on ($Y^*,X^*)$ or ($Y^*, X^*$, $Z$), and stratum sizes are determined by the adaptive grid search algorithm. This design is included as a ``gold standard'' as it requires knowledge of $\pmb{\theta}$. \textit{optMLE-2}: Subjects are jointly stratified on ($Y^*,X^*)$ or ($Y^*, X^*$, $Z$). In Phase II(a), $n / 2$ subjects are selected by BCC*. In Phase II(b), $n / 2$ more subjects are selected by the adaptive grid search algorithm, with $\pmb{\theta}$ estimated using Phase I and II(a) data.

We compared the designs using two precision measures: (i) relative efficiency (RE), defined as the ratio of empirical variances of $\hat{\beta}$ in the final analysis, and (ii) relative interquartile range (RI), defined as the ratio of the widths of the empirical interquartile range (IQR) (McIsaac \& Cook, 2015). The optMLE design based on true parameter values and observed Phase I stratum sizes $\left\{N_{y^*x^*}\right\}$ or $\left\{N_{y^*x^*z}\right\}$ was treated as the reference design when calculating the RE and RI {\color{black} (i.e., the variance and IQR, respectively, of the optMLE were used in the numerators of these measures)}. RE and RI values $>1$ indicate better precision than the optMLE design, while values $<1$ indicate worse precision. We also considered alternative versions of the referential optimal design (Supplemental Table~S1), but results were similar to the optMLE and thus they were not included in subsequent simulations.

\subsection{Outcome and exposure misclassification}\label{Paper3_Res_Both}

We simulated data for a Phase I sample of $N = \num{10000}$ subjects according to Equation \eqref{joint}. We generated $X$ and $Y$ from Bernoulli distributions with $p_{x} = P(X=1)$ and $P(Y=1|X) = [1 + \exp\{-(\beta_0 + 0.3 X)\}]^{-1}$, respectively. We used approximate outcome prevalence {\color{black}$p_{y0} = P(Y=1|X=0)$ to define $\beta_0 = \log\left\{p_{y0} / (1 - p_{y0})\right\}$}. We generated $Y^*$ and $X^*$ from Bernoulli distributions with $P(X^*=1|Y,X) = [1 + \exp\{-(\gamma_0 + 0.45 Y + \gamma_1 X)\}]^{-1}$ and $P(Y^*=1|X^*,Y,X) = [1 + \exp\{-(\alpha_0 + 0.275X^* + \alpha_1 Y + 0.275 X)\}]^{-1}$, respectively, where $(\gamma_0, \gamma_1)$ and $(\alpha_0, \alpha_1)$ control the strength of the relationship between error-prone and error-free variables. We define the ``baseline'' false positive and true positive rates for $X^*$, denoted by $\FPR_{0}(X^*)$ and $\TPR_{0}(X^*)$, respectively, as the false positive and true positive rates of $X^*$ when $Y=0$. Similarly, {\color{black}$\FPR_{00}(Y^*)$ and $\TPR_{00}(Y^*)$} are the baseline false positive and true positive rates for $Y^*$ when {\color{black}$X=0$ and $X^*=0$}. With these definitions, we have $\gamma_{0} = -\log\left\{\frac{1 - \FPR_{0}(X^*)}{\FPR_{0}(X^*)}\right\}$, $\gamma_1 = -\log\left\{\frac{1 - \TPR_{0}(X^*)}{\TPR_{0}(X^*)}\right\} - \gamma_0$, {\color{black}$\alpha_{0} = -\log\left\{\frac{1 - \FPR_{00}(Y^*)}{\FPR_{00}(Y^*)}\right\}$, and $\alpha_1 =  -\log\left\{\frac{1 - \TPR_{00}(Y^*)}{\TPR_{00}(Y^*)}\right\} - \alpha_0$}. Both $Y^*$ and $X^*$ are misclassified, but the misclassification rates are varied separately; we fix {\color{black}$\FPR(\cdot) = 0.1$ and $\TPR(\cdot) = 0.9$} for one variable and vary {\color{black}$\FPR(\cdot) = 0.1$ or $0.5$ and $\TPR(\cdot) = 0.9$ or $0.5$ for the other}. {\color{black}
We consider cases where $Y = 0$ or $1$ is more common by fixing $p_{x} = 0.1$ and varying $p_{y0}$ from $0.1$ to $0.9$. Similarly, cases where $X = 0$ or $1$ is more common were considered by fixing $p_{y0} = 0.3$ and varying $p_{x}$ from $0.1$ to $0.9$.} Using the designs in Section~\ref{Paper3_Des}, $n = 400$ subjects were selected in Phase II. We considered minimum stratum sizes of $m =$ 10--50 for the optMLE design; all yielded stable estimates (Supplemental Figure~S1), so $m = 10$ was used hereafter. {\color{black} In choosing $m$, there is a trade-off between stability of the design and potential efficiency gain, driven by larger and smaller choices of $m$, respectively.} While the grid search parameters  used to locate the optMLE and optMLE-2 designs varied between replicates, three-iteration grid searches with step sizes $\pmb{s} = \{15, 5, 1\}$ and $\pmb{s} = \{25, 5, 1\}$, respectively, were most common. Each setting was replicated \num{1000} times.

Tables~\ref{tab:YXboth_varyYerrors} and \ref{tab:YXboth_varyXerrors} show that the optMLE-2 design was highly efficient with RE $>$ 0.9 and RI $>$ 0.95 in most settings. In some settings, the RE and RI for the optMLE-2 design were even slightly larger than one; this is because the optMLE design is asymptotically optimal but may not necessarily be optimal in finite samples. In most settings, the optMLE-2 design exhibited sizeable efficiency gains over the BCC*, CC*, and SRS designs, with the gains as high as 43\%, 74\%, and 83\%, respectively. The efficiency gains were higher when the misclassification rates were lower, particularly for $Y^*$, or when Phase I stratum sizes were less balanced. {\color{black}Even under the most severe misclassification (e.g., when $\FPR_{00}(Y^*)$ and $\TPR_{00}(Y^*)$ were both $0.5$), the MLE should remain identifiable because of the validation sample. However, the CC* and SRS designs incurred bias, as much as 22\% and 27\%, respectively, primarily when $p_{y0}$ was farther from $0.5$; in these situations, imbalance in $Y$ made these designs susceptible to empty cells in the validation data. The optimal and BCC* designs were reasonably unbiased in all settings, although we saw the smallest efficiency gain for the optimal designs when the misclassification rates were highest, since the Phase I data were not very informative for Phase II.} The grid search successfully located the optMLE and optMLE-2 designs in all and $\geq 95\%$ replicates per setting, respectively.

The grid search failed to locate the optMLE-2 design in a few replicates because empty cells in the {\color{black} cross-tabulation of} unvalidated and validated data from the Phase II(a) sample {\color{black}(e.g., no ``false negatives'' for the exposure)} led to unstable coefficients {\color{black} from logistic regression} that rendered singular information matrices. {\color{black} This can happen when error rates are extreme in either direction. When error rates are extremely low, error-prone variables can become collinear with their error-free counterparts. In this case, we might treat the variable as error-free and use the Phase I version of this variable in all models. 
When error rates are extremely high, we might not observe any cases of agreement in Phase II(a) (e.g., no records with $X^* = X$). In this case, more than two waves of Phase II might be needed to ``fill in'' the empty cells.} Fortunately, we did not encounter this problem very often; out of \num{20000} total replicates across these settings, we discarded 172 (0.9\%) problematic ones.
 
Supplemental Figure~S2 shows the average Phase II stratum sizes for all designs under the settings described. The makeup of the optMLE design depended on the Phase I stratum sizes and misclassification rates. It oversampled subjects from the less-frequent strata. Furthermore, oversampling of the less-frequent strata was more extreme when misclassification rates for the variable were higher. The optMLE-2 design had similar but less extreme allocation compared to the optMLE design because it contained a BCC* sample of 200 subjects in Phase II(a). When Phase I variables were not very informative about the Phase II ones {\color{black}(e.g., $\FPR_{0}(X^*) = \TPR_{0}(X^*) = 0.5$)}, the optimal designs became less dependent on the Phase I stratum sizes, and the optMLE-2 design also became less similar to the optMLE design because estimating $\pmb{\theta}$ was harder (Supplemental Figure~S3). {\color{black} With larger minimum $m = 50$, allocation of the optMLE and optMLE-2 designs was more similar than with $m = 10$, especially when the Phase I variables were informative (Figure S4 in Web Appendix B).} 

\subsection{{\color{black}More than two waves of validation}}

{\color{black}We considered another approximately optimal design, the \textit{optMLE-3}, which conducted validation in three waves. Phase II(a) was the same as the optMLE-2, with $n/2$ subjects selected by BCC* based on ($Y^*$, $X^*$). Then, $n / 4$ subjects were selected in Phases II(b) and II(c) by the adaptive grid search algorithm, with $\btheta$ estimated from Phases I and II(a) and I, II(a), and II(b), respectively.  
Data were generated following Section~\ref{Paper3_Res_Both} with $p_{x} = 0.1$,  $p_{y0} = 0.3$, varied outcome misclassification rates, and fixed $\FPR_{0}(X^*) = 0.1$, $\TPR_{0}(X^*)=0.9$ for the exposure. Efficiency gains with the optMLE-3 were similar to the optMLE-2, with both recovering $\geq 88\%$ of the efficiency of the optMLE (Table~\ref{tab:3wave}), and the designs chose almost identical stratum sizes for validation (Supplemental Figure~S5). Other allocations of the Phase II sample across multiple validation waves are of course possible; we refer the reader to McIsaac \& Cook (2015) for additional considerations.}

\subsection{Incorporating an additional error-free covariate}\label{Paper3_Res_Both_inclZ}

We performed an additional set of simulations that incorporated an error-free covariate into the designs and analyses. Simulation details are in Web Appendix {\color{black}B.1} in the Supplementary Materials. In summary, the optMLE-2 design continued to be highly efficient, with gains as high as 43\%, 56\%, and 59\% over the BCC*, CC*, and SRS designs, respectively (Supplemental Table~S2). The optimal designs favored sampling subjects from strata with larger ${\rm Var}(X|Z=z)$, where the true value of $X$ was harder to ``guess" (Supplemental Figure~{\color{black}S5}).

\subsection{Optimal designs' robustness to model misspecification}\label{Paper3_Res_Misspec}

Next, we investigated the impact of model misspecification at the design stage on the efficiency of subsequent estimators. We simulated data using Equation~\eqref{joint} for a Phase I sample of $N = \num{10000}$ subjects. We generated an additional error-free binary covariate $Z$ along with $X$ and $Y$ from Bernoulli distributions with $P(Z = 1) = 0.25$, $P(X=1|Z) = [1 + \exp\{-(-2.2 + 0.5Z)\}]^{-1}$ and $P(Y = 1|X, Z)=[1 + \exp\{-(-0.85 + 0.3X + 0.25Z)\}]^{-1}$, respectively. We set {\color{black}$\FPR_{0}(X^*) = \FPR_{00}(Y^*) = 0.25$ and $\TPR_{0}(X^*) = \TPR_{00}(Y*) = 0.75$}, such that $X^*$ and $Y^*$ were generated from Bernoulli distributions with $P(X^*=1|Y,X,Z) = [1 + \exp\{-(-1.1 + 0.45 Y + 2.2 X + Z + \delta_{1} X Z)\}]^{-1}$ and $P(Y^*=1|X^*,Y,X,Z) = [1 + \exp\{-(-1.1 + 0.275X^* + 2.2 Y + 0.275 X + Z + \delta_{2} X Z)\}]^{-1}$, respectively, with $\delta_{1}$ and $\delta_{2}$ between {\color{black}$-1$ and $1$}. We defined eight ($Y^*$, $X^*$, $Z$) sampling strata and selected $n = 400$ subjects in Phase II. Additional optimal designs, denoted optMLE* and optMLE-2*, assumed only main effects for $P(X^*=1|Y,X,Z)$ and $P(Y^*=1|X^*,Y,X,Z)$; clearly, these models will be misspecified at the design stage for $\delta_1$ and/or $\delta_{2} \neq 0$. The analysis models were correctly specified {\color{black}(i.e., included the interaction term)}, although Lotspeich et al. (2021) found the MLE to be fairly robust to model misspecification in this setting.

Simulation results for the proposed designs can be found in Table~\ref{tab:mod_misspec}. Even though the optMLE* and optMLE-2* designs were computed based on incorrect model specifications, very little efficiency was lost relative to the correctly specified gold standard optMLE design. Moreover, the optMLE-2* design remained more efficient than existing designs, with efficiency gains as high as {\color{black}47\%, 43\%, and 37\%} over the BCC*, CC*, and SRS designs, respectively (Supplemental Table~S3). Thus, the proposed optimal designs appeared to maintain their advantages even when we were uncertain about the model specification at the design stage. {\color{black}This includes the problematic situation where, at the design stage, we incorrectly omitted an interaction term from our misclassification model.} 
The average validation sample sizes in each stratum for all designs are displayed in Supplemental Figure~{\color{black}S6}. The differences between the optMLE and optMLE* designs, or between the optMLE-2 and optMLE-2* designs, were almost always small, although more visible when the model for $Y^*$ was misspecified.

\subsection{Classical scenarios with outcome or exposure misclassification alone}\label{Paper3_Res_Classical}

Detailed results for settings with outcome or exposure misclassification only are presented in Appendices B.2 and B.3, respectively, in the Supplementary Materials. The optimal designs oversampled subjects from strata that corresponded to the less-frequent value of the error-prone variable (Supplemental Figure~{\color{black}S7 and S8}). The optMLE-2 design approximated the optMLE design well, continuing to offer sizable efficiency gains over existing designs (Supplemental Table~S4).

\section{COMPARING PARTIAL TO FULL AUDIT RESULTS IN THE VCCC}\label{Paper3_VCCC}

The VCCC EHR contains routinely collected patient data, including demographics, antiretroviral therapy (ART), labs (e.g., viral load and CD4 count), and clinical events. Since the VCCC data had been fully validated, available pre-/post-validation datasets could be used to compare two-phase designs and analyses that only validate a subset of records to the gold standard analysis that uses the fully validated data. We used these data to assess the relative odds of having an AIDS-defining event (ADE) within one year of ART initiation {\color{black}($Y$)} between patients who were/were not ART naive at enrollment {\color{black}($X$)}, adjusting for square root transformed CD4 at ART initiation {\color{black}($Z$)}. We extracted $N = 2012$ records from the EHR for this study In the unvalidated data {\color{black}($Y^*$, $X^*$)}, 73\% of patients were ART naive at enrollment and 8\% of patients experienced an ADE within one year. 

The misclassification rate of ADE was 6\%, with 63\% false positive rate (FPR) and only 1\% false negative rate (FNR). The misclassification rate of the ART naive status at enrollment was 11\%, with FPR $=13\%$ and FNR $=3\%$. Only 19 subjects (1\%) had both outcome and exposure misclassification. CD4 count was error-free. {\color{black} We assumed misclassification models for ADE and ART status, with $P_{\pmb{\alpha}}(Y^*=1|X^*,Y,X,Z) = [1 + \exp\{-(\alpha_{0} + \alpha_{1}X^* + \alpha_{2}Y + \alpha_{3}X + \alpha_{4}Z)\}]^{-1}$ and $P_{\pmb{\gamma}}(X^*=1|Y,X,Z) = [1 + \exp\{-(\gamma_{0} + \gamma_{1}Y + \gamma_{2}X + \gamma_{3}Z)\}]^{-1}$, respectively.}

We defined four strata according to the unvalidated Phase I ADE and ART naive status, with stratum sizes $(N_{00}, N_{01}, N_{10}, N_{11}) =$ (504, 1350, 42, 116), where the first and second subscripts index error-prone ADE and ART naive status, respectively. We set $n = 200$ and considered the optMLE-2, BCC*, CC*, and SRS designs. When implementing the optMLE-2 design, we selected $n^{(a)} = 100$ subjects in Phase II(a) via BCC* sampling. All results were averaged over 1000 replicates except for SRS and optMLE-2 designs: SRS encountered 118 replicates where the MLE was unstable or did not converge because of very small numbers of audited events or exposures, while the grid search algorithm failed to locate the optMLE-2 design in 40 of the replicates. On average, the SRS, CC*, BCC*, and optMLE-2 audits chose $(n_{00}, n_{01}, n_{10}, n_{11}) =$ (56, 134, 4, 12), (27, 73, 26, 74), (53, 53, 42, 52), and (25, 39, 42, 95) subjects, respectively, from the four strata in Phase II. 

Table~\ref{tab:vccc_cd4} shows the results under the two-phase designs and those from the gold standard and naive analyses using fully validated and unvalidated data, respectively, from the full cohort. The log OR estimates under all two-phase designs were reasonably close to the gold standard estimates and led to the same clinical interpretations, i.e., after controlling for $\sqrt{CD4}$ at ART initiation, ART naive status at enrollment was not associated with changes in the odds of ADE within one year of ART initiation. The variance under the optMLE-2 design was 14\%, 13\%, and 86\% smaller than those under the BCC*, CC*, and SRS designs, respectively. 

We also considered designs that further stratified on CD4 count. Specifically, we dichotomized CD4 count at its median, $238 \text{ cells/mm}^3$, and formed strata defined by the error-prone ADE, ART status, and CD4 category. The Phase I stratum sizes $(N_{000}, N_{010}, N_{100}, N_{110}$, $N_{001}, N_{011}, N_{101}$, $N_{111}) =$ (171, 701, 34, 93, 333, 649, 8, 23), where the third subscript indexed CD4 category, with 0 and 1 corresponding to low and high CD4 counts, respectively. CD4 count was still treated continuously in all analyses. The SRS and CC* designs were unchanged because they do not incorporate exposure or covariate information. The BCC* design selected (28, 28, 28, 29, 28, 28, 8, 23) subjects in Phase II. The grid search algorithm successfully located the optMLE-2 design in 952 of the replicates (95\%) with average Phase II stratum sizes (14, 32, 33, 70, 13, 13, 8, 16). This resulted in very minor efficiency gains {\color{black}over the optMLE-2 that did not stratify by CD4 category} (Table~\ref{tab:vccc_cd4}).

\section{PROSPECTIVE AUDIT PLANNING IN CCASANET}\label{Paper3_CCASAnet}

Researchers are interested in assessing the association between bacteriological confirmation of TB and successful treatment outcomes among PLWH who are treated for TB. We are in the process of designing a multi-site audit of $n = 500$ patients to validate key variables and better estimate this association in the CCASAnet cohort. The outcome of interest ($Y$) is successful completion of TB treatment within one year of diagnosis; among patients who did not complete treatment, this captures unfavorable outcomes of death, TB recurrence, or loss to follow-up (with each of these outcomes also of interest in secondary analyses). The exposure of interest ($X$) is bacterial confirmation of TB, defined as any positive diagnostic test result, e.g., culture, smear, or PCR. 

The Phase I sample comes from the current CCASAnet research database and includes all patients initiating TB treatment between January 1, 2010 and December 31, 2018. Error-prone values $(Y^*,X^*)$ of the study variables are available on $N = 3478$ TB cases across sites in five countries (anonymously labeled as Countries A--E) during this period. Patients were stratified on $(Y^*,X^*)$ within Countries A--E to create strata of sizes $(N_{00}, N_{01}, N_{10}, N_{11})$ = (704, 246, 1015, 415), (239, 139, 336, 218), (3, 7, 5, 17), (6, 9, 15, 14), and (12, 16, 36, 26), respectively.

To implement the optMLE-2 design as in Sections~\ref{Paper3_Res}--\ref{Paper3_VCCC}, $n^{(a)} = 250$ patients would be chosen in Phase II(a) using BCC* sampling from the 20 ($Y^*, X^*,$ Country) strata. Alternatively, we could incorporate prior data from on-site chart reviews conducted in the five CCASAnet sites between 2009--2010. The original data from this time period captured a total of 595 TB cases (Phase I). In this historical dataset, 70\% of cases completed treatment within one year and 68\% had bacteriological confirmation of TB. Validated TB treatment and diagnosis were available for 40 subjects who were chosen for validation via site-stratified SRS. We observed 13\% and 20\% misclassification in $Y^*$ (FPR $=7\%$, FNR $=23\%$) and $X^*$ (FPR $=39\%$, FNR $=5\%$), respectively. No subject had both their outcome and exposure misclassified.

We demonstrate two ways to use these historic audits to design a more efficient validation study for the next round of CCASAnet audits. Strategy (i) estimates the parameters with the historic data, denoted $\widehat{\pmb{\theta}}^{(h)}$, and uses them to derive the optMLE design to allocate $n = 500$ subjects in one Phase II subsample. Strategy (ii) is a multi-wave strategy that uses $\widehat{\pmb{\theta}}^{(h)}$ to design Phase II(a) and then uses the Phase II(a) parameters, denoted $\widehat{\pmb{\theta}}^{(a)}$, to design Phase II(b). 

Given the small size of the historic audit ($n = 40$), it was not possible to obtain country-level estimates of all parameters needed to derive the optimal design. Instead, we created country groupings {\color{black}($Z$)} based on site-specific audit results (Supplemental Table~S5), where {\color{black}$Z = 0$} for Countries A--B with errors in $Y^*$ or $X^*$,  {\color{black}$Z = 1$} for Countries C--D with errors in both $Y^*$ and $X^*$, and {\color{black}$Z = 2$} for Country E which had no errors. {\color{black} We assumed misclassification models for TB treatment completion and bacteriological confirmation, with $P_{\pmb{\alpha}}(Y^*=1|X^*,Y,X,Z) = (1 + \exp[-\{\alpha_{0} + \alpha_{1}X^* + \alpha_{2}Y + \alpha_{3}X + \alpha_{4}{\rm I}(Z = 1) + \alpha_{5}{\rm I}(Z = 2)\}])^{-1}$ and $P_{\pmb{\gamma}}(X^*=1|Y,X,Z) = (1 + \exp[-\{\gamma_{0} + \gamma_{1}Y + \gamma_{2}X + \gamma_{3}{\rm I}(Z = 1) + \gamma_{4}{\rm I}(Z = 2)\}])^{-1}$, respectively.} These groupings were used to obtain the MLE for the historic data, $\widehat{\pmb{\theta}}^{(h)}$. Since audits will be conducted at the site level, we applied these parameters to the 20 Phase I strata from the current data by assuming the same coefficients for countries in the same CoG group (Supplemental Table~S6).

First, we derived the optMLE design for $n = 500$ {\color{black}with $m = 10$} using the historic audits (Strategy (i)), which was made up of the following $(Y^*, X^*)$ strata from Countries A--E, respectively: $(n_{00}, n_{01}, n_{10}, n_{11})$ $=$ (10, 15, 10, 21), (20, 80, 11, 168), (3, 7, 5, 17), (6, 9, 15, 14), and (12, 16, 35, 26). All, or nearly all, available subjects were taken from Countries C--E. In Countries A and B, subjects with $X^* = 1$ were preferred, particularly when paired with $Y^* = 1$. With fewer records from Country B than A, the optMLE design selected more subjects from the former.

Then, we implemented the grid search to select $n^{(a)} = 250$ subjects as a more informed first wave for the two-wave design (Strategy (ii)). Based on the historic parameters, stratum sizes to be sampled at Phase II(a) were $(n^{(a)}_{00}, n^{(a)}_{01}, n^{(a)}_{10}, n^{(a)}_{11})$ = (10, 10, 10, 10), (10, 10, 10, 45), (3, 7, 5, 17), (6, 9, 10, 13), and (12, 16, 11, 26) for Countries A--E, respectively. Validation appeared focused on the smaller countries (C--E). Country A was sampled minimally, proportional to its Phase I sample size, as were all strata in Country B except ($Y^* = 1, X^* = 1$). Validated data on these subjects will be used to re-estimate the model parameters and derive the optimal allocation for Phase II(b). Alternatively, the Phase II(a) and historic data could be pooled to re-estimate the parameters. In our situation, the historic audits were much smaller than the Phase II(a) study, so Phase II(a) would likely dominate the pooled analysis. However, if the Phase II(a) study were smaller, i.e., due to budget constraints, then the benefits of data pooling could be greater.

Ultimately, the choice between these strategies is determined by logistics and our confidence in the historic data. We plan to use Strategy (ii): the optMLE-2 design with the first wave informed by prior audits. Incorporating the prior audit information, even if it might be biased, will likely be better than starting off with a BCC* design (Chen \& Lumley, 2020), 
but we do not want {\color{black}to trust the historic audits entirely}. Also, conducting multiple validation waves is feasible because they can be performed by in-country investigators (Lotspeich et al., 2020).

\section{DISCUSSION}\label{Paper3_Discuss}

\begin{figure}
\includegraphics[width=\linewidth]{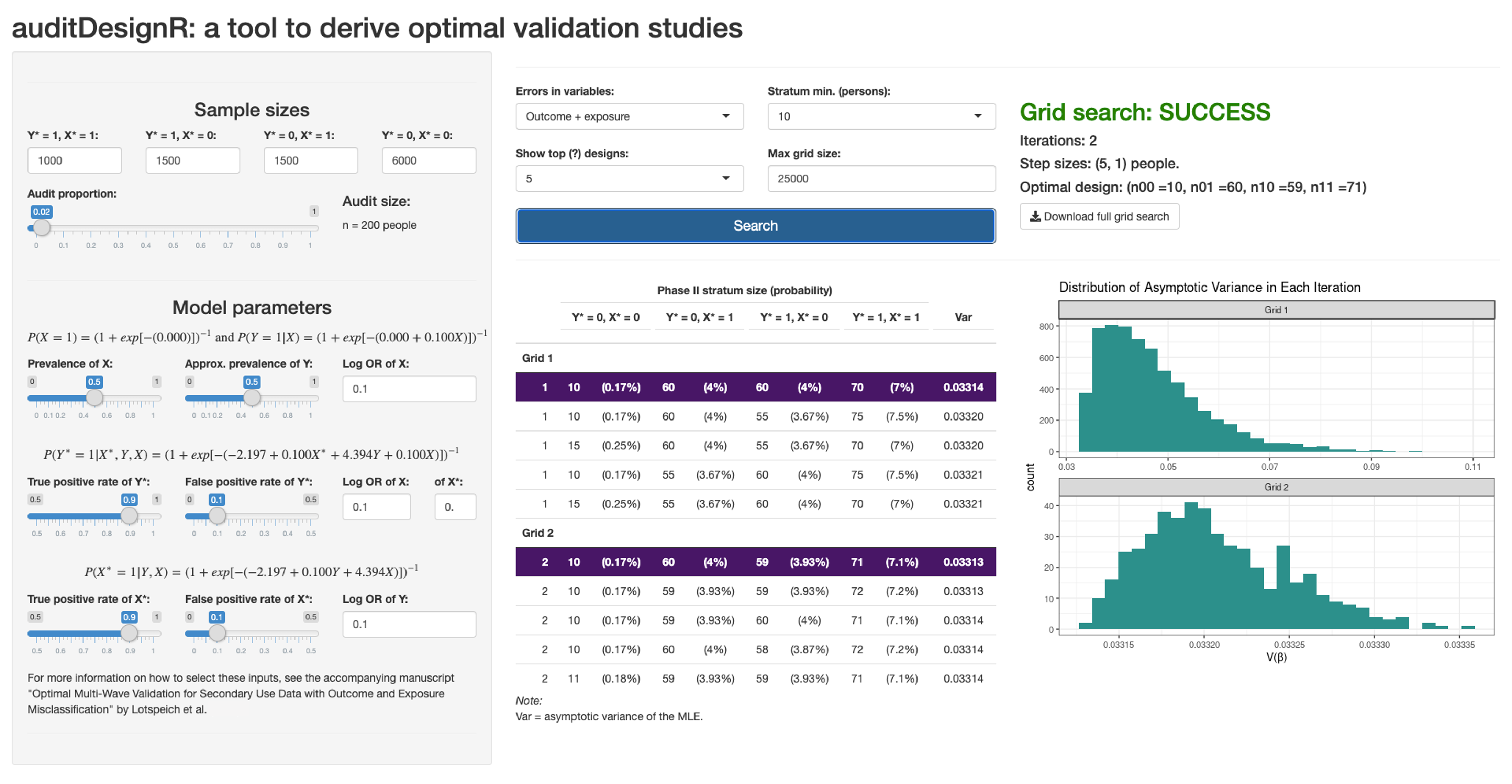}
\caption{A screenshot of the \textit{auditDesignR} Shiny application after a $T = 2$ iteration adaptive grid search to find the optimal design. The user selects options from the grey sidebar (e.g., Phase I strata sizes and model parameters), followed by design-specific options like which error setting is assumed (e.g., errors in both ``Outcome + exposure'') and the minimum required stratum size $m$. There are also controls for the adaptive grid search routine, like the maximum allowable grid size (here assumed to be \num{25000} candidate designs). After making their selections and clicking ``Search,'' the user can view the top candidate designs in addition to the distribution of ${\rm Var}(\hat{\beta})$ from each iteration of the grid search.}
\label{shiny}
\end{figure}

Validation studies are integral to many statistical methods to correct for errors in secondary use data. However, they are resource-intensive undertakings. The number of records and variables that can be reviewed are limited by time, budget, and staffing constraints. Thus, selecting the most informative records is key to maximizing the benefits of validation. We introduced a new optimal design, and a multi-wave approximation to it, which maximizes the efficiency of the MLE under differential outcome and exposure misclassification -- a setting for which optimal designs have not yet been proposed. {\color{black}We devise a novel adaptive grid search algorithm to locate the optimal design, and the designs are implemented in the \textit{auditDesignR} R package and Shiny app (Supporting Information, Figure~\ref{shiny}).}

{\color{black}As part of our audit protocol, the CCASAnet sites are notified in advance with the list of patient records to be validated. This provides time for site investigators to locate the relevant patient charts before the audit, but there is still a chance that validation data may be missing. Our methods rely on the MAR assumption, which asserts that conditioning on the Phase I information, subjects who are unaudited are similar to those who are audited. When we select who to audit, MAR holds because the validation data are missing by design, but when validation data are missing simply because we cannot find them, it calls the MAR assumption into question. In our previous CCASAnet audits, there have been instances where validation data are missing for selected records, although it is not very common. In the past, we have implicitly assumed that these records are MAR in analyses. Methods to simultaneously address audit data that are missing by design (i.e., MAR) and nonresponse (i.e., possibly not MAR) would be an interesting direction for future work.}

Our analyses and designs are based on the parametric MLE approach of Tang et al. (2015). Recently, an SMLE approach was developed to analyze two-phase studies with error-prone outcome and exposure data that nonparametrically models the exposure error mechanism, making it robust to different exposure error mechanisms (Lotspeich et al., 2021). 
Our designs are guaranteed to be optimal for the MLE but still offer efficiency gains for the SMLE; this avoids complicated calculations which would be required to derive a design specifically for the SMLE. In an additional simulation, we found that the efficiency gains of the SMLE and MLE under the proposed optimal designs were essentially identical (Supplemental Table~S7). 

{\color{black} We focused on designs for full-likelihood estimators because they offer the greatest efficiency if full-cohort information (i.e., both Phases I and II data) are available, which was the case with the VCCC and CCASAnet data examples. However, if only audit data are available or one wants to avoid placing additional models on the misclassification mechanisms, conditional maximum likelihood estimation (CMLE) could be considered (Breslow \& Cain, 1988). Optimal designs for CMLE could be calculated in a similar manner, but we have not done so here.}

Strictly speaking, the proposed optimal design is ``optimal" among designs with compatible strata definitions only. {\color{black}In the VCCC example, we considered an optimal design that sampled on a categorical version of continuous CD4 count in addition to the error-prone outcome and exposure; we created two categories based on CD4 counts above or below the median. While we treated CD4 as discrete to compute the optimal design, it still performed well with continuous CD4 in the analysis. There are other ways we might have discretized CD4, e.g., by creating more than two categories or choosing a different cutpoint than the median. How to best stratify continuous variables for design purposes is an interesting question (e.g., Amorim et al., 2021), and designs that maintain the continuous nature of continuous covariates warrant further investigation.}

Other interesting topics of future research include developing optimal designs for two-phase validation studies with other types of outcomes and exposures, including count, continuous, or censored data. {\color{black} Developing these designs would involve replacing the models in Equations~\eqref{joint}--\eqref{I_theta_theta} with appropriate ones for the new data types and then deriving the successive steps in parallel to the current work. The special case with Phase I data made up of multiple error-prone surrogates for $Y$ or $X$ (e.g., $X_{1}^*, X_2^*, \dots, X_{p}^*$), as in a reliability study, would also be a natural extension of the methods herein. Also, the proposed designs could be modified to pursue optimal estimation of the interaction between the misclassified exposure and an additional covariate, similar to the focus of Breslow \& Chatterjee (1999), or of multiple parameters simultaneously (e.g., the main effects and interaction). Either of these modifications would require adoption of an alternative criterion to summarize the variance matrix, such as D-optimality or A-optimality which minimize the determinant or trace, respectively, of the variance matrix (Fedorov \& Leonov, 2013).}

\section*{APPENDIX}

\subsection*{Derivations of $S^{v}(\cdot)$ and $S^{\bar{v}}(\cdot)$}\label{AppC_Deriv}

{\color{black}Recall that Equation~\eqref{joint} is the joint density of a complete observation, which includes the validation indicator along with error-prone and error-free variables. Now, the joint density of just the error-prone and error-free variables, needed to derive the score vectors, is defined $P(Y^*,X^*,Y,X,\pmb{Z}) =$
\begin{align*}
& P(Y^*|X^*,Y,X,\pmb{Z})P(X^*|Y,X,\pmb{Z})P(Y|X,\pmb{Z})P(X|\pmb{Z})P(\pmb{Z}).
\end{align*}
Then, depending on whether $V_i=1$ or 0 ($i = 1, \dots, N$), we denote the log-likelihood contribution of the $i$th subject by $l^v(\pmb{\theta};Y_i^*,X_i^*,Y_i,X_i,\pmb{Z}_i) = \log P_{\btheta}(Y_i^*,X_i^*,Y_i,X_i,\pmb{Z}_i)$ or $l^{\bar{v}}(\pmb{\theta};Y_i^*,X_i^*,\pmb{Z}_i) = \log P_{\btheta}(Y_i^*,X_i^*,\pmb{Z}_i) = \log\left\{\sum_{y=0}^{1}\sum_{x=0}^{1} P_{\btheta}(Y_i^*,X_i^*,y,x,\pmb{Z}_i)\right\}$, respectively. Recognize that $l^v(\pmb{\theta};Y_i^*,X_i^*,Y_i,X_i,\pmb{Z}_i)$ and $l^{\bar{v}}(\pmb{\theta};Y_i^*,X_i^*,\pmb{Z}_i)$ are equivalent to the summands in the first and second terms of the observed-data log-likelihood from Equation~\eqref{od_ll}.} 

Now, we denote the score vector for the $i$th subject {\color{black}based on their log-likelihood contribution} by $S_i(\pmb{\theta}) = (S_i(\beta), S_i(\pmb{\eta})^{\rm T})^{\rm T}$, where
\begin{align*}
S_i(\theta_j) =& V_i\frac{\partial}{\partial\theta_j}l^v(\pmb{\theta};Y_i^*,X_i^*,Y_i,X_i,\pmb{Z}_i)+(1-V_i)\frac{\partial}{\partial\theta_j}l^{\bar{v}}(\pmb{\theta};Y_i^*,X_i^*,\pmb{Z}_i)\nonumber\\ 
\equiv&V_iS^v(\theta_j; Y_i^*,X_i^*,Y_i,X_i,\pmb{Z}_i) + (1-V_i) S^{\bar{v}}(\theta_j;Y_i^*,X_i^*,\pmb{Z}_i),\ \forall\ \theta_j \in \pmb{\theta}. 
\end{align*}
We decompose $\pmb{\eta}^{\rm T}$ into $(\pmb{\eta}_{y^*}^{\rm T}, \pmb{\eta}_{x^*}^{\rm T}, \pmb{\eta}_{y}^{\rm T}, \pmb{\eta}_{x}^{\rm T})^{\rm T}$, where $\pmb{\eta}_{y^*}$, $\pmb{\eta}_{x^*}$, $\pmb{\eta}_{y}$, and $\pmb{\eta}_{x}$ correspond to the nuisance parameters in models $P_{\pmb{\eta}_{y^*}}(Y^*|X^*,Y,X,\pmb{Z})$, $P_{\pmb{\eta}_{x^*}}(X^*|Y,X,\pmb{Z})$, $P_{\beta,\pmb{\eta}_{y}}(Y|X,\pmb{Z})$, and $P_{\pmb{\eta}_{x}}(X|\pmb{Z})$, respectively. Then, we {\color{black} derive the score vectors as} 
\begin{align}
&S^{v}(\theta_j;Y_i^*,X_i^*,Y_i,X_i,\pmb{Z}_i) \nonumber \\
=& \begin{cases} 
\left\{\frac{\partial}{\partial\theta_j}P_{\pmb{\eta}_{y^*}}(Y_i^*|X_i^*, Y_i, X_i,\pmb{Z}_i)\right\}\left\{P_{\pmb{\eta}_{y^*}}(Y_i^*|X_i^*, Y_i, X_i,\pmb{Z}_i)\right\}^{-1}, & \text{ if }\theta_j \in \pmb{\eta}_{y^*}, \\
\left\{\frac{\partial}{\partial\theta_j}P_{\pmb{\eta}_{x^*}}(X_i^*|Y_i, X_i,\pmb{Z}_i)\right\}\left\{P_{\pmb{\eta}_{x^*}}(X_i^*|Y_i, X_i, \pmb{Z}_i)\right\}^{-1}, & \text{ if }\theta_j \in \pmb{\eta}_{x^*}, \\
\left\{\frac{\partial}{\partial\theta_j}P_{\beta,\pmb{\eta}_{y}}(Y_i|X_i,\pmb{Z}_i)\right\}\left\{P_{\beta,\pmb{\eta}_{y}}(Y_i|X_i, \pmb{Z}_i)\right\}^{-1}, & \text{ if }\theta_j \in (\beta, \pmb{\eta}_{y}), \\
\left\{\frac{\partial}{\partial\theta_j}P_{\pmb{\eta}_{x}}(X_i|\pmb{Z}_i)\right\}\left\{P_{\pmb{\eta}_{x}}(X_i|\pmb{Z}_i)\right\}^{-1}, & \text{ if }\theta_j \in \pmb{\eta}_{x},
\end{cases} \nonumber\\
& S^{\bar{v}}(\theta_j;Y_i^*,X_i^*,\pmb{Z}_i) \nonumber \\
= & \frac{\sum_{y=0}^{1}\sum_{x=0}^{1}\frac{\partial}{\partial\theta_j}\left\{P_{\pmb{\eta}_{y^*}}(Y_i^*|X_i^*,y,x,\pmb{Z}_i)P_{\pmb{\eta}_{x^*}}(X_i^*|y,x,\pmb{Z}_i)P_{\beta,\pmb{\eta}_{y}}(y|x,\pmb{Z}_i)P_{\pmb{\eta}_{x}}(x|\pmb{Z}_i)\right\}}{\sum_{y=0}^{1}\sum_{x=0}^{1}P_{\pmb{\eta}_{y^*}}(Y_i^*|X_i^*,y,x,\pmb{Z}_i)P_{\pmb{\eta}_{x^*}}(X_i^*|y,x,\pmb{Z}_i)P_{\beta,\pmb{\eta}_{y}}(y|x,\pmb{Z}_i)P_{\pmb{\eta}_{x}}(x|\pmb{Z}_i)}. \nonumber
\end{align}

\clearpage

\section*{FIGURES AND TABLES}

\begin{table}[h]
\caption{\label{tab:YXboth_varyYerrors}Simulation results under outcome and exposure misclassification with varied outcome misclassification rates.}%
\resizebox{\textwidth}{!}{
\begin{tabular}{cclrccclrccclrccclrccc}
\toprule
\multicolumn{2}{c}{\textbf{Outcome Misclassification}} & \multicolumn{1}{c}{\textbf{ }} & \multicolumn{4}{c}{\textbf{optMLE-2}} & \multicolumn{1}{c}{\textbf{ }} & \multicolumn{4}{c}{\textbf{BCC*}} & \multicolumn{1}{c}{\textbf{ }} & \multicolumn{4}{c}{\textbf{CC*}} & \multicolumn{1}{c}{\textbf{ }} & \multicolumn{4}{c}{\textbf{SRS}} \\
\textbf{$\pmb{\FPR_{00}(Y^*)}$} & \textbf{$\pmb{\TPR_{00}(Y^*)}$} && \textbf{\% Bias} & \textbf{SE} & \textbf{RE} & \textbf{RI} && \textbf{\% Bias} & \textbf{SE} & \textbf{RE} & \textbf{RI} && \textbf{\% Bias} & \textbf{SE} & \textbf{RE} & \textbf{RI} && \textbf{\% Bias} & \textbf{SE} & \textbf{RE} & \textbf{RI} \\
\hline
\multicolumn{22}{c}{$p_{y0} = 0.1$} \\
$0.1$ & $0.9$ &  & $-2.85$ & $0.214$ & $1.028$ & $1.015$ &  & $-4.00$ & $0.254$ & $0.728$ & $0.855$ &  & $ -9.20$ & $0.347$ & $0.391$ & $0.640$ &  & $-19.4$ & $0.516$ & $0.176$ & $0.459$\\
& $0.5$ &  & $1.11$ & $0.241$ & $0.908$ & $0.960$ &  & $ 1.39$ & $0.286$ & $0.643$ & $0.815$ &  & $ -3.98$ & $0.362$ & $0.403$ & $0.679$ &  & $-17.2$ & $0.512$ & $0.201$ & $0.508$\\
$0.5$ & $0.9$ &  & $-2.99$ & $0.321$ & $0.935$ & $1.017$ &  & $-3.08$ & $0.409$ & $0.578$ & $0.763$ &  & $-21.6$ & $0.560$ & $0.308$ & $0.570$ &  & $-17.7$ & $0.563$ & $0.305$ & $0.552$\\
& $0.5$ &  & $-1.61$ & $0.361$ & $1.067$ & $1.008$ &  & $-6.31$ & $0.377$ & $0.982$ & $1.004$ &  & $-10.8$ & $0.512$ & $0.531$ & $0.767$ &  & $-27.0$ & $0.543$ & $0.472$ & $0.700$\\
\multicolumn{22}{c}{$p_{y0} = 0.3$} \\
$0.1$ & $0.9$ &  & $1.67$ & $0.190$ & $1.009$ & $0.949$ &  & $0.34$ & $0.223$ & $0.734$ & $0.855$ &  & $ 7.26$ & $0.297$ & $0.413$ & $0.683$ &  & $-1.46$ & $0.333$ & $0.329$ & $0.569$\\
& $0.5$ &  & $4.83$ & $0.219$ & $1.004$ & $1.049$ &  & $-0.75$ & $0.226$ & $0.941$ & $1.087$ &  & $6.54$ & $0.317$ & $0.480$ & $0.723$ &  & $1.38$ & $0.344$ & $0.406$ & $0.658$\\
$0.5$ & $0.9$ &  & $0.62$ & $0.241$ & $0.924$ & $0.879$ &  & $-0.27$ & $0.274$ & $0.717$ & $0.814$ &  & $-7.60$ & $0.386$ & $0.360$ & $0.588$ &  & $2.02$ & $0.357$ & $0.421$ & $0.626$\\
& $0.5$ &  & $1.73$ & $0.248$ & $0.918$ & $0.957$ &  & $-0.81$ & $0.240$ & $0.982$ & $1.035$ &  & $-1.15$ & $0.369$ & $0.416$ & $0.664$ &  & $-1.01$ & $0.369$ & $0.415$ & $0.675$\\
\multicolumn{22}{c}{$p_{y0} = 0.75$} \\
$0.1$ & $0.9$ && $1.87$ & $0.206$ &  $0.917$ & $0.950$ && $0.88$ & $0.227$ & $0.833$ & $0.840$ && $4.11$ & $0.360$ & $0.525$ & $0.585$ && $9.61$ & $0.405$ & $0.467$ & $0.532$ \\
& $0.5$ && $2.44$ & $0.273$ & $0.978$ & $0.960$ && $3.79$ & $0.303$ & $0.881$ & $0.887$ && $15.8$ & $0.473$ & $0.564$ & $0.573$ && $11.2$ & $0.439$ & $0.608$ & $0.628$ \\
$0.5$ & $0.9$ && $-3.28$ & $0.248$ & $0.940$ & $0.907$ && $-0.05$ & $0.267$ & $0.873$ & $0.872$ && $0.09$ & $0.390$ & $0.597$ & $0.601$ && $24.4$ & $0.451$ & $0.517$ & $0.530$ \\
& $0.5$ && $8.79$ & $0.301$ & $1.010$ & $1.030$ && $4.04$ & $0.313$ & $0.974$ & $0.905$ && $6.89$ & $0.465$ & $0.656$ & $0.601$ && $7.05$ & $0.445$ & $0.685$ & $0.699$ \\
\hline
\end{tabular}
}
\noindent {\scriptsize {\color{black}Exposure misclassification rates were fixed at $\FPR_{0}(X^*) = 0.1$, $\TPR_{0}(X^*) = 0.9$.} \% Bias and SE are, respectively, the empirical {\color{black}percent} bias and standard error of the MLE. {\color{black}RE or RI $< 1$ indicates an efficiency loss compared to the optMLE.} The grid search successfully located the optMLE and optMLE-2 designs in all and $\geq 95\%$ replicates per setting, respectively; across all settings, 162 ($1.4$\%) problematic replicates of the optMLE-2 were discarded out of \num{12000}. Fewer than $1\%$ and $5\%$ of the replicates were discarded because of unstable estimates under the SRS, CC*, or BCC* designs when $p_{y0} = 0.1$ and $0.9$, respectively. All other entries are based on \num{1000} replicates.}
\end{table}

\begin{table}[h]
\caption{\label{tab:YXboth_varyXerrors}Simulation results under outcome and exposure misclassification with varied exposure misclassification rates.}%
\resizebox{\textwidth}{!}{
\begin{tabular}{cclrccclrccclrccclrccc}
\toprule
\multicolumn{2}{c}{\textbf{Exposure Misclassification}} & \multicolumn{1}{c}{\textbf{ }} & \multicolumn{4}{c}{\textbf{optMLE-2}} & \multicolumn{1}{c}{\textbf{ }} & \multicolumn{4}{c}{\textbf{BCC*}} & \multicolumn{1}{c}{\textbf{ }} & \multicolumn{4}{c}{\textbf{CC*}} & \multicolumn{1}{c}{\textbf{ }} & \multicolumn{4}{c}{\textbf{SRS}} \\
\textbf{$\pmb{\FPR_{0}(X^*)}$} & \textbf{$\pmb{\TPR_{0}(X^*)}$} && \textbf{\% Bias} & \textbf{SE} & \textbf{RE} & \textbf{RI} && \textbf{\% Bias} & \textbf{SE} & \textbf{RE} & \textbf{RI} && \textbf{\% Bias} & \textbf{SE} & \textbf{RE} & \textbf{RI} && \textbf{\% Bias} & \textbf{SE} & \textbf{RE} & \textbf{RI} \\
\hline
\multicolumn{22}{c}{$p_{x} = 0.1$} \\
$0.1$ & $0.9$ &  & $1.67$ & $0.190$ & $1.009$ & $0.949$ &  & $ 0.34$ & $0.223$ & $0.734$ & $0.855$ &  & $7.27$ & $0.297$ & $0.413$ & $0.683$ &  & $-1.46$ & $0.333$ & $0.329$ & $0.569$\\
& $0.5$ &  & $4.17$ & $0.218$ & $1.000$ & $0.990$ &  & $ 0.92$ & $0.247$ & $0.781$ & $0.860$ &  & $3.39$ & $0.336$ & $0.420$ & $0.600$ &  & $-1.54$ & $0.338$ & $0.414$ & $0.637$\\
$0.5$ & $0.9$ &  & $1.74$ & $0.296$ & $0.973$ & $1.015$ &  & $-2.07$ & $0.351$ & $0.691$ & $0.866$ &  & $3.79$ & $0.342$ & $0.730$ & $0.885$ &  & $-0.06$ & $0.351$ & $0.693$ & $0.866$\\
& $0.5$ &  & $2.75$ & $0.343$ & $1.020$ & $1.027$ &  & $ 0.19$ & $0.342$ & $1.028$ & $0.993$ &  & $2.45$ & $0.348$ & $0.993$ & $1.026$ &  & $ 4.72$ & $0.357$ & $0.942$ & $0.997$\\
\multicolumn{22}{c}{$p_{x} = 0.9$} \\
$0.1$ & $0.9$ &  & $2.08$ & $0.189$ & $0.851$ & $0.940$ &  & $-1.14$ & $0.201$ & $0.750$ & $0.879$ &  & $3.88$ & $0.310$ & $0.316$ & $0.584$ &  & $0.08$ & $0.339$ & $0.265$ & $0.520$\\
& $0.5$ &  & $-5.23$ & $0.290$ & $0.960$ & $0.910$ &  & $4.23$ & $0.343$ & $0.685$ & $0.811$ &  & $2.37$ & $0.345$ & $0.678$ & $0.771$ &  & $0.89$ & $0.381$ & $0.555$ & $0.750$\\
$0.5$ & $0.9$ &  & $-2.00$ & $0.221$ & $0.977$ & $0.920$ &  & $ 2.64$ & $0.264$ & $0.681$ & $0.838$ &  & $-2.07$ & $0.337$ & $0.418$ & $0.600$ &  & $5.52$ & $0.366$ & $0.355$ & $0.576$\\
& $0.5$ &  & $5.88$ & $0.364$ & $1.008$ & $0.984$ &  & $7.80$ & $0.366$ & $0.996$ & $0.975$ &  & $2.12$ & $0.360$ & $1.029$ & $0.983$ &  & $12.0$ & $0.387$ & $0.890$ & $0.941$\\
\hline
\end{tabular}}
\noindent {\scriptsize  {\color{black}Outcome misclassification rates were fixed at $\FPR_{00}(Y^*) = 0.1$, $\TPR_{00}(Y^*) = 0.9$.} \% Bias and SE are, respectively, the empirical percent bias and standard error of the MLE. {\color{black}RE or RI $< 1$ indicates an efficiency loss compared to the optMLE.} The grid search successfully located the optMLE and optMLE-2 designs in all and $\geq 99\%$ replicates per setting, respectively; across all settings, 10 ($< 0.1$\%) problematic replicates of the optMLE-2 were discarded out of \num{8000}. All other entries are based on \num{1000} replicates.}
\end{table}

\begin{table}[h]
\caption{\label{tab:3wave}\color{black}Simulation results with two- or three-wave approximate optimal designs under outcome and exposure misclassification.}%
\resizebox{\textwidth}{!}{
\begin{tabular}{cclrccclrccc}
\toprule
\multicolumn{2}{c}{\textbf{Outcome Misclassification}} && \multicolumn{4}{c}{\textbf{optMLE-2}} && \multicolumn{4}{c}{\textbf{optMLE-3}}  \\
\textbf{$\pmb{\FPR_{00}(Y^*)}$} & \textbf{$\pmb{\TPR_{00}(Y^*)}$} && \textbf{\% Bias} & \textbf{SE} & \textbf{RE} & \textbf{RI} && \textbf{\%Bias} & \textbf{SE} & \textbf{RE} & \textbf{RI} \\
\hline
$0.1$ & $0.9$ && $1.33$ & $0.190$ & $1.009$ & $0.949$ &&  $3.33$ & $0.191$ & $1.000$ & $0.959$ \\
& $0.5$ && $5.00$ & $0.219$ & $1.004$ & $1.049$ &&  $1.33$ & $0.217$ & $1.028$ & $1.060$ \\
$0.5$ & $0.9$ && $0.67$ & $0.241$ & $0.924$ & $0.879$ && $-2.00$ & $0.235$ & $0.975$ & $0.879$ \\
& $0.5$ && $1.67$ & $0.248$ & $0.918$ & $0.957$ && $2.00$ & $0.251$ & $0.899$ & $0.943$ \\
\hline
\end{tabular}
}
\noindent {\scriptsize {\color{black} Outcome misclassification rates were varied and exposure misclassification rates were fixed at $\FPR_{0}(X^*) = 0.1$, $\TPR_{0}(X^*) = 0.9$.} \% Bias and SE are, respectively, the empirical bias and standard error of the MLE. {\color{black}RE or RI $< 1$ indicates an efficiency loss compared to the optMLE.} The grid search successfully located the optMLE, optMLE-2, and optMLE-3 designs in all, $\geq 99\%$, and $\geq 98\%$ of replicates per setting, respectively; across all settings, 24 ($0.6\%$) and 43 ($1.1\%$) problematic replicates of the optMLE-2 and optMLE-3, respectively, were discarded out of \num{4000}. All other entries are based on \num{1000} replicates.}
\end{table}

\begin{table}[h]
\caption{\label{tab:mod_misspec}Simulation results when the {\color{black}misspecification models} used in the optimal design can be misspecified.}%
\resizebox{\linewidth}{!}{
\begin{tabular}{rrlrccccrcccrrccc}
\toprule
\multicolumn{3}{c}{} & 
\multicolumn{4}{c}{\textbf{optMLE*}} & \multicolumn{1}{c}{} & \multicolumn{4}{c}{\textbf{optMLE-2}} & \multicolumn{1}{c}{} &  \multicolumn{4}{c}{\textbf{optMLE-2*}} \\
$\pmb{\delta_1}$ & $\pmb{\delta_2}$ && 
\textbf{\%Bias} & \textbf{SE} & \textbf{RE} & \textbf{RI} && 
\textbf{\%Bias} & \textbf{SE} & \textbf{RE} & \textbf{RI} && 
\textbf{\%Bias} & \textbf{SE} & \textbf{RE} & \textbf{RI}  \\
\hline
\multicolumn{17}{c}{\textbf{Misspecified misclassification mechanism for $\pmb{Y^*}$ and $\pmb{X^*}$}} \\
$-1.0$ & $-1.0$ &  & $ 2.13$ & $0.286$ & $0.927$ & $0.985$ &  & $ 0.42$ & $0.290$ & $0.903$ & $0.914$ &  & $ 2.61$ & $0.280$ & $0.964$ & $1.011$\\
$-0.5$ & $-0.5$ &  & $ 6.20$ & $0.252$ & $1.025$ & $1.008$ &  & $-1.24$ & $0.262$ & $0.949$ & $0.999$ &  & $-1.44$ & $0.261$ & $0.956$ & $0.970$\\
$ 0.0$ & $ 0.0$ &  & $-2.35$ & $0.260$ & $1.000$ & $1.000$ &  & $ 0.17$ & $0.270$ & $0.932$ & $1.065$ &  & $ 0.17$ & $0.270$ & $0.932$ & $1.065$\\
$ 0.5$ & $ 0.5$ &  & $ 1.76$ & $0.251$ & $1.053$ & $0.974$ &  & $ 2.87$ & $0.256$ & $1.014$ & $0.971$ &  & $ 1.36$ & $0.260$ & $0.986$ & $0.978$\\
$ 1.0$ & $ 1.0$ &  & $ 2.70$ & $0.255$ & $0.944$ & $0.918$ &  & $ 4.12$ & $0.263$ & $0.883$ & $0.963$ &  & $ 5.06$ & $0.256$ & $0.936$ & $0.938$\\
\multicolumn{17}{c}{\textbf{Misspecified misclassification mechanism for $\pmb{Y^*}$}} \\
$0.0$ & $-1.0$ &  & $ 0.16$ & $0.266$ & $0.991$ & $0.911$ &  & $-5.54$ & $0.287$ & $0.855$ & $0.897$ &  & $ 0.29$ & $0.272$ & $0.948$ & $0.935$\\
$0.0$ & $-0.5$ &  & $ 4.76$ & $0.305$ & $0.865$ & $0.889$ &  & $-3.22$ & $0.296$ & $0.918$ & $0.962$ &  & $-3.74$ & $0.295$ & $0.922$ & $0.962$\\
$0.0$ & $ 0.0$ &  & $-2.35$ & $0.260$ & $1.000$ & $1.000$ &  & $ 0.17$ & $0.270$ & $0.932$ & $1.065$ &  & $ 0.17$ & $0.270$ & $0.932$ & $1.065$\\
$0.0$ & $ 0.5$ &  & $ 7.51$ & $0.279$ & $1.267$ & $1.159$ &  & $ 5.16$ & $0.305$ & $1.058$ & $1.051$ &  & $ 3.64$ & $0.301$ & $1.087$ & $1.038$\\
$0.0$ & $ 1.0$ &  & $ 4.76$ & $0.257$ & $1.018$ & $0.988$ &  & $ 0.35$ & $0.266$ & $0.952$ & $0.966$ &  & $-1.65$ & $0.258$ & $1.011$ & $0.976$\\
\multicolumn{17}{c}{\textbf{Misspecified misclassification mechanism for $\pmb{X^*}$}} \\
$-1.0$ & $0.0$ &  & $-2.22$ & $0.270$ & $1.030$ & $1.031$ &  & $-8.65$ & $0.266$ & $1.059$ & $1.001$ &  & $-8.47$ & $0.273$ & $1.005$ & $0.984$\\
$-0.5$ & $0.0$ &  & $-1.32$ & $0.252$ & $0.994$ & $0.947$ &  & $-1.12$ & $0.266$ & $0.897$ & $0.910$ &  & $-3.60$ & $0.262$ & $0.924$ & $0.999$\\
$ 0.0$ & $0.0$ &  & $-2.35$ & $0.260$ & $1.000$ & $1.000$ &  & $ 0.17$ & $0.270$ & $0.932$ & $1.065$ &  & $ 0.17$ & $0.270$ & $0.932$ & $1.065$\\
$ 0.5$ & $0.0$ &  & $-0.25$ & $0.256$ & $0.984$ & $0.982$ &  & $ 1.11$ & $0.255$ & $0.990$ & $1.022$ &  & $ 3.64$ & $0.266$ & $0.908$ & $0.908$\\
$ 1.0$ & $0.0$ &  & $-4.66$ & $0.255$ & $1.046$ & $1.035$ &  & $-0.30$ & $0.256$ & $1.037$ & $1.027$ &  & $-6.65$ & $0.258$ & $1.019$ & $0.994$\\
\hline
\end{tabular}
}
\noindent {\scriptsize {\color{black}Misclassification rates were fixed at
$\FPR(\cdot) = 0.1$, $\TPR(\cdot) = 0.9$.} The error-prone exposure and outcome were generated from models including the interaction terms $\delta_1 X Z$ and $\delta_{2} X Z$, respectively, but the optMLE* and optMLE-2* designs assumed only main effects for these models. \% Bias and SE are, respectively, the empirical bias and standard error of the MLE. {\color{black}RE or RI $< 1$ indicates an efficiency loss compared to the optMLE.} The optMLE and optMLE* were located in all replicates and the optMLE-2 and optMLE-2* designs were located in $\geq 98\%$ and $99\%$ of replicates per setting, respectively; {\color{black}65 ($ 0.5\%$) and 17 ($0.1\%$) problematic replicates out of \num{13000} were discarded for the optMLE-2 and optMLE-2*, respectively.} All other entries are based on \num{1000} replicates.}
\end{table}

\begin{table}[h]
\caption{\label{tab:vccc_cd4}Estimates and standard errors from the analysis of the VCCC dataset under the optMLE-2, BCC*, CC*, and SRS validation designs.}%
\resizebox{\linewidth}{!}{
\begin{tabular}{rlrclrclrc}
\toprule
& &  \multicolumn{2}{c}{\textbf{(Intercept)}} & & \multicolumn{2}{c}{\textbf{ART Status}} & & \multicolumn{2}{c}{$\pmb{\sqrt{CD4}}$} \\
\textbf{Design} & & \textbf{log Odds} & \textbf{SE} && \textbf{log OR} & \textbf{SE} && \textbf{log OR} & \textbf{SE} \\
\hline
\multicolumn{10}{l}{\textit{Full cohort analyses}}\\
\hspace{1em}Gold standard && $-1.184$ & $0.294$ && $0.032$ & $0.260$ && $-0.180$ & $0.022$ \\
\hspace{1em}Naive && $-0.043$ & $0.234$ &  & $-0.308$ & $0.200$ && $-0.148$ & $0.015$ \\
\multicolumn{10}{l}{\textit{Two-phase analyses}}\\
\multicolumn{10}{c}{Sampling Strata Defined by ADE and ART Status}\\
\hspace{1em}SRS  && $-1.050$ & $0.821$ && $-0.161$ & $0.996$ && $-0.189$ & $0.058$ \\
\hspace{1em}CC* && $-1.527$ & $0.438$ && $0.090$ & $0.394$ && $-0.149$ & $0.034$ \\
\hspace{1em}BCC* && $-1.325$ & $0.420$ && $0.006$ & $0.396$ && $-0.169$ & $0.035$ \\
\hspace{1em}optMLE-2  && $-1.542$ & $0.394$ && $0.118$ & $0.368$ && $-0.151$ & $0.035$ \\
\multicolumn{10}{c}{Sampling Strata Defined by ADE, ART Status, and CD4 Count}\\
\hspace{1em}BCC* && $-1.402$ & $0.421$ && $0.115$ & $0.406$ && $-0.163$ & $0.034$ \\
\hspace{1em}optMLE-2 &&  $-1.495$ & $0.392$ && $0.105$ & $0.362$ && $-0.150$ & $0.034$ \\
\hline
\end{tabular}}
\noindent {\scriptsize All results were averaged over 1000 replicates except for SRS and optMLE-2 designs: SRS encountered 118 replicates where the MLE was unstable or did not converge because of very small numbers of audited events or exposures, while the grid search algorithm failed to locate the optMLE-2 design in 40 and 48 of the replicates, respectively, under the first and second definitions of sampling strata.}
\end{table}

\clearpage

\section*{SUPPLEMENTARY MATERIALS}

\begin{itemize}
    \item \textbf{Additional appendices, tables, and figures:} The Web Appendices and Supplemental Figures and Tables referenced in Sections 2--5 can be found in the Supplementary Materials that follow this text.
    \item \textbf{R-package for optimal designs:} R-package \textit{auditDesignR} that implements the new designs can be found at \url{https://github.com/sarahlotspeich/auditDesignR/}.
    \item \textbf{R Shiny application for optimal designs:} A Shiny app that computes the optimal design is at \url{http://ec2-54-226-78-254.compute-1.amazonaws.com/auditDesignR/}
    \item \textbf{R code for simulation studies:} R scripts to replicate the simulation studies from Section~3 can be found at \url{https://github.com/sarahlotspeich/auditDesignR/}.
\end{itemize}

\section*{ACKNOWLEDGEMENTS}
This research was supported by the Patient-Centered Outcomes Research Institute grant R-1609-36207; the National Institutes of Health grants R01AI131771, R01HL094786, and U01AI069923; and the National Institute of Environmental Health Sciences grant T32ES007018. The authors thank the VCCC and CCASAnet for permission to present their data. This work used resources of the Advanced Computing Center for Research and Education at Vanderbilt University and Longleaf at the University of North Carolina at Chapel Hill.

\section*{REFERENCES}
\begin{itemize}[leftmargin=*]
\item[]
\noindent Amorim, G., Tao, R., Lotspeich, S., Shaw, P.~A., Lumley, T. \& Shepherd, B.~E. (2021). Two-phase sampling designs for data validation in settings with covariate measurement error and continuous outcome. {\it Journal of the Royal Statistical Society. Series A, (Statistics in Society)}, 184(4), 1368--1389.

\item[]
Breslow, N.~E. \& Cain, K.~C. (1988). Logistic regression for two-stage case-control data. {\it Biometrika}, 75(1), 11--20.

\item[]
Breslow, N.~E. \& Chatterjee, N. (1999). Design and analysis of two-phase studies with binary outcome applied to Wilms tumour prognosis. {\it Journal of the Royal Statistical Society. Series C (Applied Statistics)}, 48(4), 457--468.

\item[]
Chen, T. \& Lumley, T. (2020). Optimal multiwave sampling for regression modeling in two-phase designs. {\it Statistics in Medicine}, 39(30), 4912--4921.

\item[]
Crabtree-Ramirez, B.~E., Jenkins, C., Jayathilake, K., Carriquiry, G., Veloso, V.~G., Padgett, D., Gotuzzo, E., Cortes, C., Mejia, F., McGowan, C.~C., Duda, S., Shepherd, B.~E. \& Sterling, T.~R. (2019). HIV-related tuberculosis: mortality risk in persons without vs. with culture-confirmed disease. {\it International Journal of Tuberculosis and Lung Disease}, 23, 306--314.

\item[]
Deville, J.~C., Sarndal, C.~E. \& Sautory, O. (1993). Generalized raking procedures in survey sampling. {\it Journal of the American Statistical Association}, 88(423), 1013--1020.

\item[]
Duda, S.~N., Shepherd, B.~E., Gadd, C.~S., Masys, D.~R. \& McGowan, C.~C. (2012). Measuring the quality of observational study data in an international HIV research network. {\it PloS One}, 7(4), e33908.

\item[]
{\color{black}Fedorov, V.~V. \& Leonov, S.~L. (2013). {\it Optimal Design for Nonlinear Response Models}, CRC Press, Boca Raton.}

\item[]
Giganti, M.~J., Shaw, P.~A., Chen, G., Bebawy, S.~S., Turner, M.~M., Sterling, T.~R. \& Shepherd, B.~E. (2020). Accounting for dependent errors in predictors and time-to-event outcomes using electronic health records, validation samples, and multiple imputation. {\it Annals of Applied Statistics}, 14(2), 1045--1061.

\item[]
Giganti, M.~J., Shepherd, B.~E., Caro-Vega, Y.,  Luz, P.~M., Rebeiro, P.~F. and Maia, M., Julmiste, G., Cortes, C., McGowan, C.~C. \& Duda, S.~N. (2019). The impact of data quality and source data verification on epidemiologic inference: a practical application using HIV observational data. {\it BMC Public Health}, 19(1), 1748.

\item[]
Han, K., Lumley, T., Shepherd, B.~E. \& Shaw, P.~A. (2020). Two-phase analysis and study design for survival models with error-prone exposures. {\it Statistical Methods in Medical Research}, 30(3), 857--874.

\item[]
Holcroft, C.~A. \&  Spiegelman, D. (1999). Design of validation studies for estimating the odds ratio of exposure-disease relationships when exposure is misclassified. {\it Biometrics}, 55, 1193--1201.

\item[]
Horvitz, D.~G. \& Thompson, J.~D. (1952). A generalization of sampling without replacement from a finite universe. {\it Journal of the American Statistical Association}, 47(260), 663--685.

\item[]
Keogh, R.~H., Shaw, P.~A., Gustafson, P., Carroll, R.~J., Deffner, V., Dodd, K.~W., Kuchenhoff, H., Tooze, J.~A., Wallace, M.~P., Kipnis, V. \& Freedman, L.~S. (2020). STRATOS guidance document on measurement error and misclassification of variables in observational epidemiology: Part 1—Basic theory and simple methods of adjustment. {\it Statistics in Medicine}, 39(16), 2197--2231.

\item[]
Lawless, J.~F., Kalbfleisch, J.~D. \& Wild, C.~J. (1999). Semiparametric Methods for Response-Selective and Missing Data Problems in Regression. {\it Journal of the Royal Statistical Society. Series B (Statistical Methodology)}, 61(2), 413--438.

\item[]
Lotspeich, S.~C., Giganti, M.~J., Maia, M., Vieira, R., Machado, D.~M., Succi, R.~C., Ribeiro, S., Pereira, M.~S., Rodriguez, M.~F., Julmiste, G., Luque, M.~T., Caro-Vega, Y., Mejia, F., Shepherd, B.~E., McGowan, C.~C. \& Duda, S.~N. (2020). Self-audits as alternatives to travel-audits for improving data quality in the Caribbean, Central and South America network for HIV epidemiology. {\it Journal of Clinical and Translational Science}, 4(2), 125--132.

\item[]
Lotspeich, S.~C., Shepherd, B.~E., Amorim, G.~G.~C., Shaw, P.~A. \& Tao, R. (2021). Efficient odds ratio estimation under two-phase sampling using error-prone data from a multi-national HIV research cohort. {\it Biometrics}, 1--12. https://doi.org/10.1111/biom.13512

\item[]
McGowan, C.~C., Cahn, P., Gotuzzo, E., Padgett, D., Pape, J.~W., Wolff, M., Schechter, M. \& Masys, D.~R. (2007). Cohort Profile: Caribbean, Central and South America Network for HIV research (CCASAnet) collaboration within the International Epidemiologic Databases to Evaluate AIDS (IeDEA) programme. {\it International Journal of Epidemiology}, 36(5), 969--976.

\item[]
McIsaac, M.~A. \& Cook, R.~J. (2014). Response-dependent two-phase sampling designs for biomarker studies. {\it The Canadian Journal of Statistics}, 42(2), 268--284.

\item[]
McIsaac, M.~A. \& Cook, R.~J. (2015). Adaptive sampling in two-phase designs: A biomarker study for progression in arthritis. {\it Statistics in Medicine}, 34, 2899--2912.

\item[]
Oh, E.~J., Shepherd, B.~E., Lumley, T. \& Shaw, P.~A. (2021a). Improved generalized raking estimators to address dependent covariate and failure-time outcome error. {\it Biometrical Journal}, 63(5), 1006--1027.

\item[]
Oh, E.~J., Shepherd, B.~E., Lumley, T. \& Shaw, P.~A. (2021b). Raking and regression calibration: Methods to address bias from correlated covariate and time-to-event error. {\it Statistics in Medicine}, 40(3), 631--649.

\item[]
Reilly, M. \& Pepe, M.~S. (1995). A mean score method for missing and auxiliary covariate data in regression models. {\it Biometrika}, 82(2), 299--314.

\item[]
Robins, J.~M., Rotnitzky, A. \& Zhao, L.~P. (1994). Estimation of regression coefficients when some regressors are not always observed. {\it Journal of the American Statistical Association}, 89(427), 846--866.

\item[]
Safran, C., Bloomsrosen, M., Hammond, E., Labkoff, S., Markel-Fox, S., Tang, P.~C., \& Detmer, D.~E. (2007). Toward a national framework for the secondary use of health data: An American Medical Informatics Association White Paper. {\it Journal of the American Medical Informatics Association}, 14(1), 1--9.

\item[]
{\color{black}Shepherd, B.~E., Han, K., Chen, T., Bian, A., Pugh, S., Duda, S.~N., Lumley, T., Heerman, W.~J., \& Shaw, P.~A. (2022) Multiwave validation sampling for error-prone electronic health records. {\it Biometrics}, 00, 1--15. https://doi.org/10.1111/biom.13713}

\item[]
Tang, L., Lyles, R.~H., King, C.~C., Celentano, D.~D. \& Lo, Y. (2015). Binary regression with differentially misclassified response and exposure variables. {\it Statistics in Medicine}, 34(9), 1605--1620.

\item[]
{\color{black}Tan, W.~K. \& Heagerty, P.~J. (2015). Surrogate-guided sampling designs for classification of rare outcomes from electronic medical records data. {\it Biostatistics}, 23(2), 345--361.}

\item[]
Tao, R., Zeng, D. \& Lin, D.~Y.(2020). Optimal designs of two-phase studies. {\it Journal of the American Statistical Association}, 115(532), 1946--1959.

\item[]
{\color{black}Wang, L., Damrauer, S.~M., Zhang, H., Zhang, A.~X., Xiao, R., Moore, J.~H. \& Chen, J. (2017). Phenotype validation in electronic health records based genetic association studies. {\it Genetic Epidemiology}, 41(8), 790--800.}

\item[]
White, J.~E. (1982). A two stage design for the study of the relationship between a rare exposure and a rare disease. {\it American Journal of Epidemiology}, 115(1), 119--128.

\item[]
{\color{black}Zhou, H., Weaver, M.~A., Qin, J., Longnecker, M.~P. \& Wang, M.~C. (2002). A semiparametric empirical likelihood method for data from an outcome-dependent sampling scheme with a continuous outcome. {\it Biometrics}, 58, 413--421.}
\end{itemize}


\end{document}


\maketitle

\section{{Choosing Step Sizes for the Adaptive Grid Search}}

As stated in the text, our adaptive grid search algorithm locates the optimal design by searching a series of grids, which are ``adaptively'' constructed at iteratively finer scales and over more focused candidate design spaces. The choice of the step sizes (or scales) of the grids appears inconsequential, but we detail the implementation used by \textit{auditDesignR} to suggest them. Our software assumes a user-specified maximum allowable grid size, which would be dictated by their machine; we use \num{10000} as the maximum for Sections 3--5.

We want to choose the first step size, $s^{(1)}$, to be the largest value for which the dimension of $\pmb{G}^{(1)}$ still falls below the allowed maximum. Calculating the dimension of a grid based on a potential step size, $s$, involves applying the ``stars and bars'' problem from combinatorics. Based on the audit size constraint (Equation~(5) in the text), the number of ``stars'' to partition is equal to $(n - Km) / s$, the number of subjects to allocate (after the minimum $m$ has been dispensed to each stratum) in increments of the step size $s$, and there are $(K - 1)$ ``bars'' (i.e., partitions to form) between them. Thus, the number of rows in the first grid based on a step size of $s$ is 
\begin{align}
rows\left\{\pmb{G}^{(1)}|s\right\} &= \binom{(n-Km)/s + (K - 1)}{(K-1)}.\label{stars&bars}    
\end{align}
For simplicity, we start by considering all possible values $s$ that share common factors. (This also ensures overlap between the neighborhoods in successive iterations such that, as a safety net, no candidate designs are ``left out''.) In the example from Section~2.3, we have $n = 400$, $K = 4$, and $m = 10$, we consider possible step sizes $\pmb{s} = \{180, 90, 45, 15, 5, 1\}$, which lead to possible grids with $rows\left\{\pmb{G}^{(1)}|\pmb{s}\right\} = \{10, 35, 165, 2925, 67525, 7906261\}$, respectively, following Equation~\eqref{stars&bars}. Thus, we select $s^{(1)} = 15$, since it is the largest potential step size in $\pmb{s}$ to keep the grid smaller than \num{10000} rows. 

There is slightly more to consider in choosing step sizes $s^{(t)}$ for successive iterations $t > 1$. We still want to cover the entire candidate design space in an efficient way. On top of that, the candidate design space has now narrowed to be in the $s^{(t-1)}$-person window around the last iteration's ``best'' design, so based on $\{n^{(t-1)}_{y^*x^*\pmb{z}}\}$ we want to impose lower- and upper-bounds on the stratum sizes considered. Calculating the size of a grid based on a possible step size of $s$ where $n_{y^*x^*z}^{(t)} \geq (n^{(t-1)}_{y^*x^*\pmb{z}} - s^{(t-1)})$ (i.e., all candidate designs are above the lower-bound of the neighborhood) involves a 
modification to Equation~\eqref{stars&bars}:
\begin{align}
& rows\left\{\pmb{G}^{(t)}|s,s^{(t-1)},n^{(t)}_{y^*x^*\pmb{z}}\geq n^{(t-1)}_{y^*x^*\pmb{z}}\right\} \nonumber \\
&= \binom{(n-Km)/s + (K - 1) - \sum_{y^*=0}^{1}\sum_{x^*=0}^{1}\sum_{\bz}(n^{(t-1)}_{y^*x^*\pmb{z}} - s^{(t-1)})/s}{(K-1)}.\label{stars&bars-lb}
\end{align}
Still, we need to subtract from Equation~\eqref{stars&bars-lb} the number of candidate designs where the stratum sizes are above the upper-bound of the neighborhood. In \textit{auditDesignR} we manually tabulate the number of such designs and subtract it from Equation~\eqref{stars&bars-lb} to calculate the expected grid size for $s$ in iteration $t$, $rows\left\{\pmb{G}^{(t)}|s,s^{(t-1)}, n^{(t-1)}_{y^*x^*\pmb{z}}\right\}$. As before, we consider possible values in $\pmb{s}$ that share common factors, but now we also want to also focus on $\pmb{s} < s^{(t-1)}$. In the second iteration of the example from Section~2.3, we consider $\pmb{s} = \{5, 1\}$, which were expected to lead to grids with $rows\left\{\pmb{G}^{(t)}|\pmb{s},s^{(t-1)}, n^{(t-1)}_{y^*x^*\pmb{z}}\right\} = \{134, 10296\}$, respectively. Thus we select $s^{(2)} = 5$, since it is the largest (and only) step size considered that keeps the grid smaller than \num{10000} rows. This process is repeated until we can reasonably reach a step size of $s^{(T)} = 1$ while keeping the size of the grid below the maximum.

\clearpage
\section{Additional Simulations, Tables, and Figures}\label{AppC_AddlSims}

\begin{table}[!h]
\caption{\label{tab:gold_stand}Three versions of the optimal design under outcome and exposure misclassification.}
\begin{center}
{\tabcolsep=4.25pt
\resizebox{\textwidth}{!}{
\begin{tabular}{@{}cclrclrccclrccc@{}}
\hline 
\multicolumn{15}{c}{\textbf{(a) Exposure misclassification rates were fixed at FPR$\pmb{_{0}(X^*) = 0.1}$ and FPR$\pmb{_{0}(X^*) = 0.9}$}}\\
\multicolumn{2}{c}{\textbf{Outcome Misclassification}} & \multicolumn{1}{c}{} &  \multicolumn{2}{c}{\textbf{\textbf{optMLE}}} & \multicolumn{1}{c}{} & \multicolumn{4}{c}{\textbf{optMLE-EXP}} &  \multicolumn{1}{c}{} & \multicolumn{4}{c}{\textbf{optMLE-FC}} \\
\cmidrule(l{3pt}r{3pt}){1-2}\cmidrule(l{3pt}r{3pt}){4-5} \cmidrule(l{3pt}r{3pt}){7-10} \cmidrule(l{3pt}r{3pt}){12-15}
\textbf{$\pmb{\FPR_{00}(Y^*)}$} & \textbf{$\pmb{\TPR_{00}(Y^*)}$} && \textbf{\% Bias} & \textbf{SE} && \textbf{\% Bias} & \textbf{SE} & \textbf{RE} & \textbf{RI} && \textbf{\% Bias} & \textbf{SE} & \textbf{RE} & \textbf{RI} \\
\hline
$0.1$ & $0.9$ &  & $-1.999$ & $0.191$ &  & $-2.033$ & $0.193$ & $0.977$ & $0.983$ &  & $0.198$ & $0.192$ & $0.987$ & $0.976$\\
& $0.5$ &  & $ 1.020$ & $0.220$ &  & $ 4.607$ & $0.225$ & $0.956$ & $1.019$ &  & $2.340$ & $0.215$ & $1.039$ & $1.012$\\
$0.5$ & $0.9$ &  & $ 1.500$ & $0.232$ &  & $ 2.659$ & $0.221$ & $1.094$ & $0.940$ &  & $2.183$ & $0.222$ & $1.093$ & $1.024$\\
& $0.5$ &  & $ 4.594$ & $0.238$ &  & $ 3.191$ & $0.239$ & $0.988$ & $0.949$ &  & $6.363$ & $0.248$ & $0.921$ & $0.982$\\
\addlinespace[0.5em]
\multicolumn{15}{c}{\textbf{(b) Outcome misclassification rates were fixed at FPR$\pmb{_{00}(Y^*) = 0.1}$ and FPR$\pmb{_{00}(Y^*) = 0.9}$}}\\
\addlinespace[0.5em]
\multicolumn{2}{c}{\textbf{Exposure Misclassification}} & \multicolumn{1}{c}{} &  \multicolumn{2}{c}{\textbf{\textbf{optMLE}}} & \multicolumn{1}{c}{} & \multicolumn{4}{c}{\textbf{optMLE-EXP}} &  \multicolumn{1}{c}{} & \multicolumn{4}{c}{\textbf{optMLE-FC}} \\
\cmidrule(l{3pt}r{3pt}){1-2}\cmidrule(l{3pt}r{3pt}){4-5} \cmidrule(l{3pt}r{3pt}){7-10} \cmidrule(l{3pt}r{3pt}){12-15}
\textbf{$\pmb{\FPR_{0}(X^*)}$} & \textbf{$\pmb{\TPR_{0}(X^*)}$} && \textbf{\% Bias} & \textbf{SE} && \textbf{\% Bias} & \textbf{SE} & \textbf{RE} & \textbf{RI} && \textbf{\% Bias} & \textbf{SE} & \textbf{RE} & \textbf{RI} \\
\hline \\
$0.1$ & $0.9$ &  & $-1.999$ & $0.191$ &  & $-2.033$ & $0.193$ & $0.977$ & $0.983$ &  & $0.198$ & $0.192$ & $0.987$ & $0.976$\\
& $0.5$ &  & $ 5.078$ & $0.218$ &  & $ 4.227$ & $0.213$ & $1.048$ & $1.002$ &  & $8.684$ & $0.216$ & $1.013$ & $0.967$\\
$0.5$ & $0.9$ &  & $ 4.508$ & $0.292$ &  & $ 8.276$ & $0.287$ & $1.033$ & $1.032$ &  & $4.243$ & $0.285$ & $1.051$ & $0.995$\\
& $0.5$ &  & $ 2.824$ & $0.347$ &  & $ 8.117$ & $0.341$ & $1.031$ & $1.007$ &  & $0.796$ & $0.338$ & $1.054$ & $1.057$\\
\hline
\end{tabular}}}    
\end{center}
{\noindent The optMLE design was based on the true parameters $\pmb{\theta}$ and the observed stratum sizes $\{N_{y^*x^*}\}$ in each replicate. The optMLE-EXP design was based on the true parameters $\pmb{\theta}$ and the expected stratum sizes ${\rm E}(N_{y^*x^*}) = N  P(Y^*=y^*,X^*=x^*)$; this design was the same for each replicate. The optMLE-FC design was based on the full cohort parameter estimates $\hat{\pmb{\theta}}$ and the observed stratum sizes $\{N_{y^*x^*}\}$ in each replicate. \% Bias and SE are, respectively, the empirical percent bias and standard error of the MLE. Each entry is based on \num{1000} replicates.
\par}
\end{table}

\begin{figure}[!h]
\centering
\includegraphics[width=0.95\textwidth]{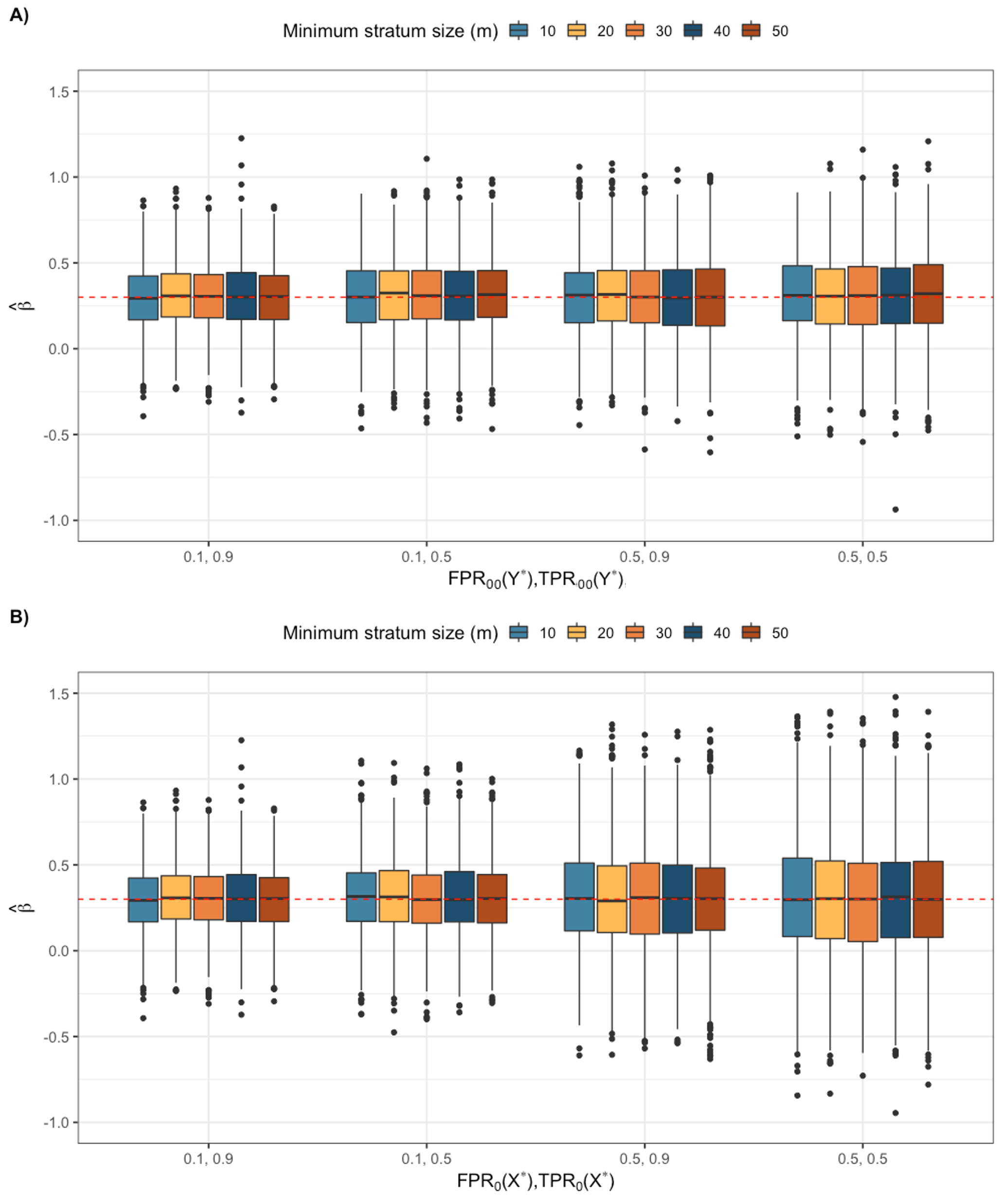}
\caption{Distribution of $\hat{\beta}$ under the optMLE design with outcome and exposure misclassification. The dashed line denotes the true value $\beta = 0.3$. Exposure and outcome misclassification rates were fixed at $\FPR_{0}(X^*) = 0.1$, $\TPR_{0}(X^*) = 0.9$ and $\FPR_{00}(Y^*) = 0.1$, $\TPR_{00}(Y^*) = 0.9$ in \textbf{A)} and \textbf{B)}, respectively.}
\label{fig:dist_est_bothYX}
\end{figure}

\begin{figure}
\centering
\includegraphics[width=\textwidth]{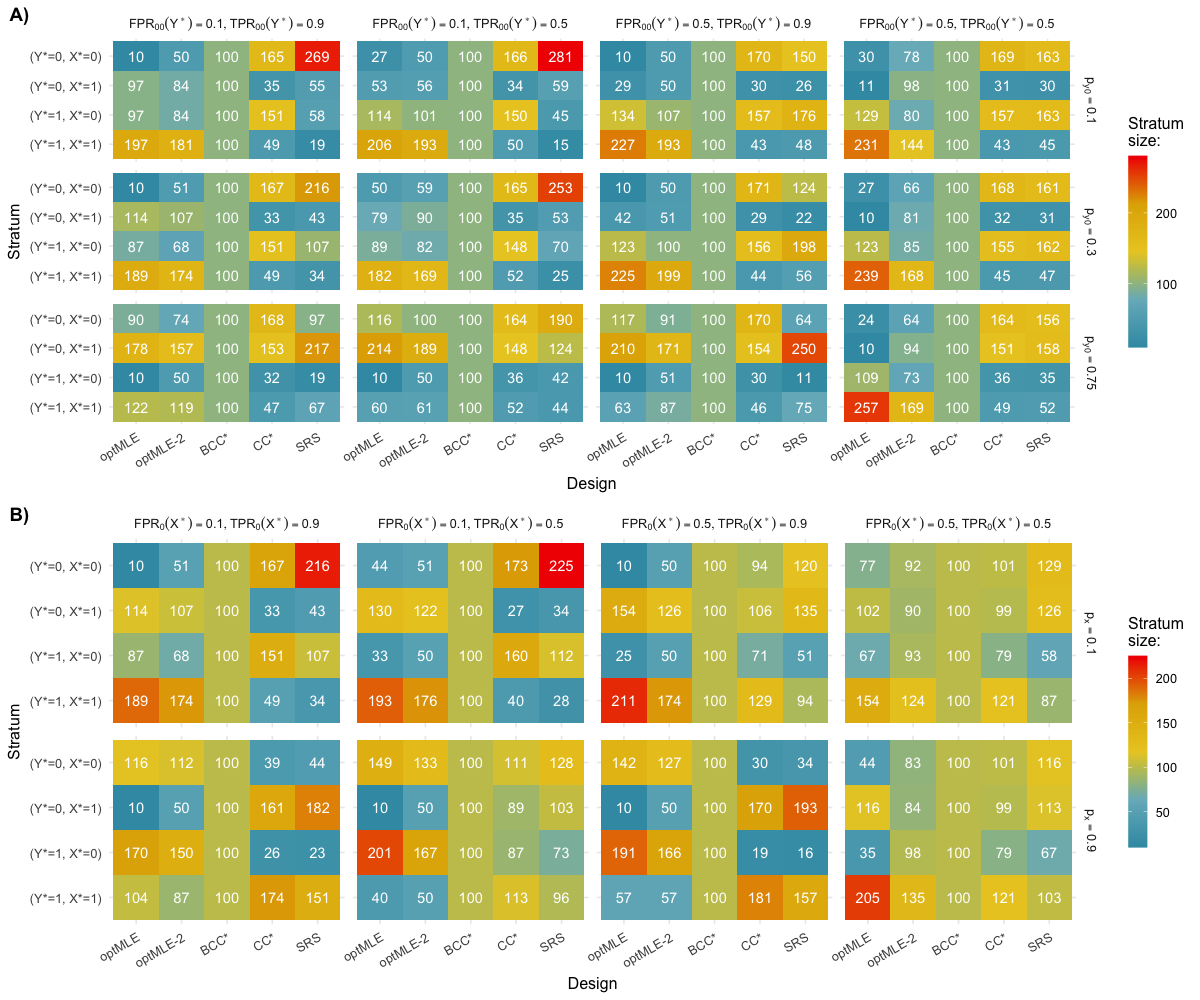}
\caption{Average Phase II stratum sizes $n_{y^*x^*}$ under outcome and exposure misclassification. Exposure and outcome misclassification rates were fixed at $\FPR_{0}(X^*) = 0.1$, $\TPR_{0}(X^*) = 0.9$ and $\FPR_{00}(Y^*) = 0.1$, $\TPR_{00}(Y^*) = 0.9$ in \textbf{A)} and \textbf{B)}, respectively.}
\label{fig:errors_YX_both}
\end{figure}

\begin{figure}
\centering
\includegraphics[width=\textwidth]{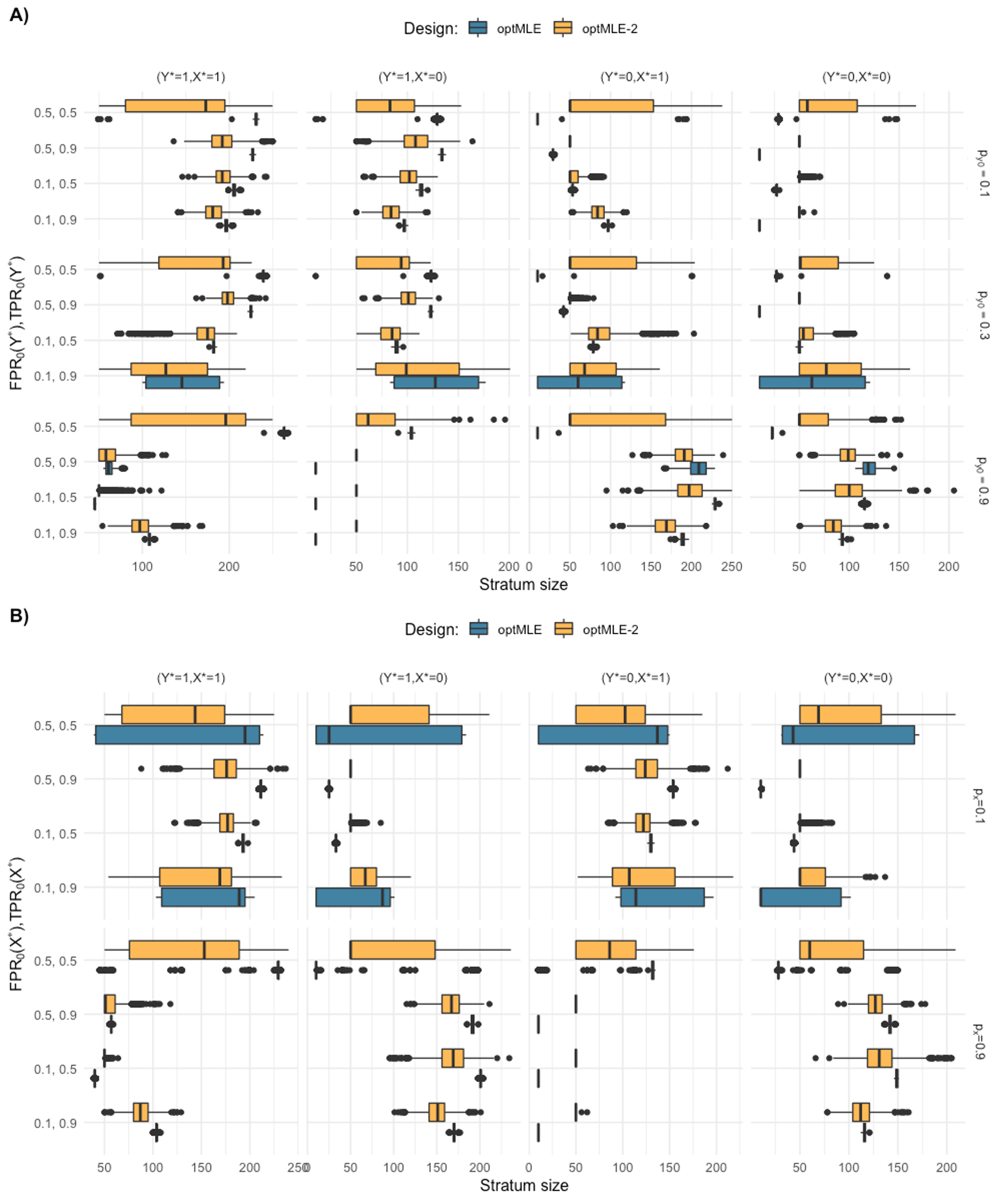}
\caption{Distribution of Phase II stratum sizes $n_{y^*x^*}$ under outcome and exposure misclassification. Exposure and outcome misclassification rates were fixed at $\FPR_{0}(X^*) = 0.1$, $\TPR_{0}(X^*) = 0.9$ and $\FPR_{00}(Y^*) = 0.1$, $\TPR_{00}(Y^*) = 0.9$ in \textbf{A)} and \textbf{B)}, respectively.}
\label{fig:dist_strat_bothYX}
\end{figure}

\begin{figure}
\centering
\includegraphics[width=\textwidth]{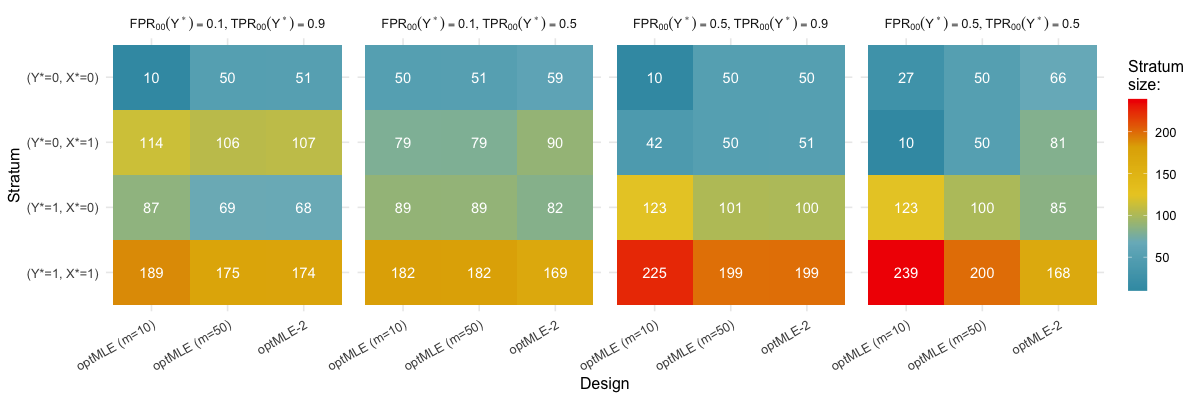}
\caption{Distribution of Phase II stratum sizes $n_{y^*x^*}$ under outcome and exposure misclassification. Two versions of the optMLE design were considered (requiring minimum stratum sizes of $m = 10$ or $50$), alongside the two-wave approximate optMLE-2 design. Exposure misclassification rates were fixed at $\FPR_{0}(X^*) = 0.1$, $\TPR_{0}(X^*) = 0.9$.}
\label{fig:optMLE_bigger_m}
\end{figure}

\begin{figure}
\centering
\includegraphics[width=\textwidth]{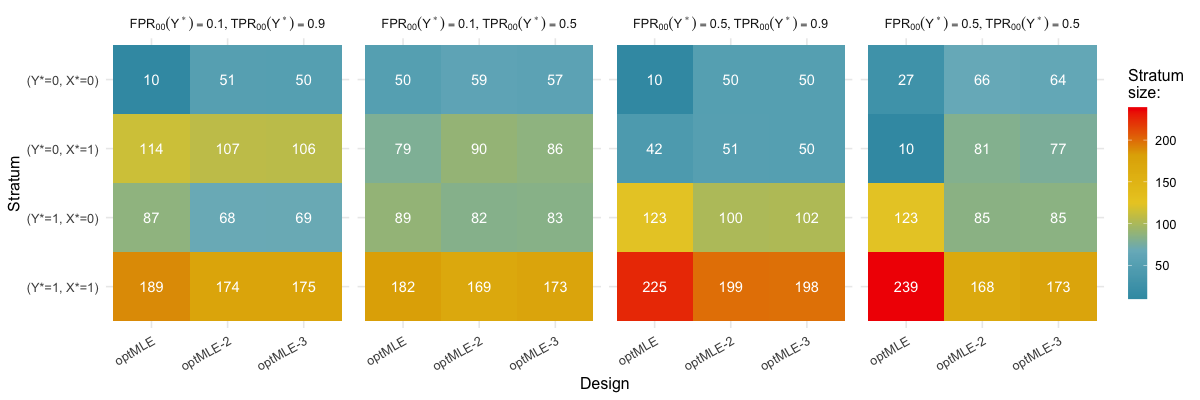}
\caption{Distribution of Phase II stratum sizes $n_{y^*x^*}$ under outcome and exposure misclassification. The optMLE design was compared to two-wave and three-wave approximations to it (the optMLE-2 and optMLE-3 designs, respectively). Exposure misclassification rates were fixed at $\FPR_{0}(X^*) = 0.1$, $\TPR_{0}(X^*) = 0.9$.}
\end{figure}

\clearpage

\subsection{Incorporating an Additional Error-Free Covariate}\label{Paper3_Res_Both_inclZ}

We simulated data using Equation~(2.1) for a Phase I sample of $N = \num{10000}$ subjects. We generated an error-free binary covariate $Z$ from a Bernoulli distribution with $P(Z = 1) \equiv p_{z} = 0.25$ or $0.5$. We generated $X$ and $Y$ from Bernoulli distributions with $P(X=1|Z) = [1 + \exp\{-(-2.2 + 0.5Z)\}]^{-1}$ and $P(Y = 1|X, Z)=[1 + \exp\{-(-0.85 + 0.3X + \beta_z Z)\}]^{-1}$, respectively, with $\beta_z=-0.25$, $0$, or $0.25$. We set the misclassification rates at $\FPR_{0}(X^*) = \FPR_{00}(Y^*) = 0.25$ and $\TPR_{0}(X^*) = \TPR_{00}(Y^*) = 0.75$, such that $X^*$ and $Y^*$ were generated from Bernoulli distributions with $P(X^*=1|Y,X,Z) = [1 + \exp\{-(-1.1 + 0.45 Y + 2.2 X + \lambda Z)\}]^{-1}$ and $P(Y^*=1|X^*,Y,X,Z) = [1 + \exp\{-(-1.1 + 0.275X^* + 2.2 Y + 0.275 X + \lambda Z)\}]^{-1}$, respectively, with $\lambda = -1$, $0$, or $1$. We defined eight sampling strata based on ($Y^*$, $X^*$, $Z$) and selected $n = 400$ subjects in Phase II. { As in Section~3.2, we set $m = 10$, but it is worth noting that larger choices of $m$ can quickly eat into our efficiency gain since we have twice as many strata. (In fact, $m = 50$ forces the optimal designs to be BCC*.)} The grid search parameters varied between replicates, but the most common choices were a six-iteration grid search with step sizes $\pmb{s} = \{40, 20, 10, 5, 2, 1\}$ and a five-iteration grid search with step sizes $\pmb{s} = \{25, 8, 5, 2, 1\}$ to locate the optMLE and optMLE-2 designs, respectively.

\begin{figure}[h]
    \centering
	   \centering
	   \includegraphics[width=\textwidth]{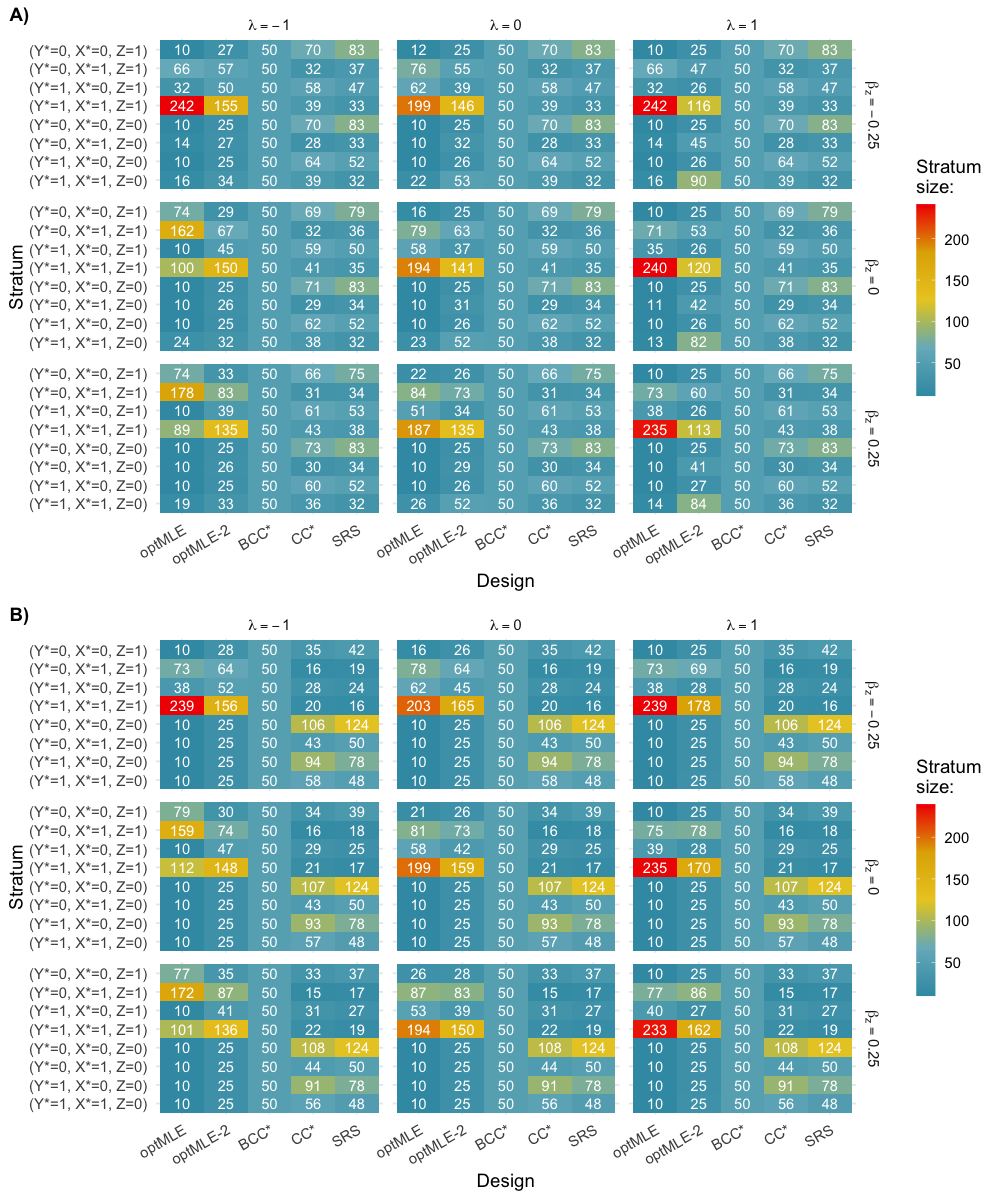}
    \caption{Average Phase II stratum sizes $n_{y^*x^*z}$ under outcome and exposure misclassification when an error-free binary covariate $Z$ with \textbf{A)} 25\% or \textbf{B)} 50\% prevalence was used in sampling.}
    \label{fig:addl_des_inclZ}
\end{figure}

Table~\ref{tab:YXinclZ} shows simulation results for the MLE under these designs. The optMLE-2 design continued to be highly efficient, with gains as high as 43\%, 56\%, and 59\% over the BCC*, CC*, and SRS designs, respectively. Fig.~\ref{fig:addl_des_inclZ} shows the average Phase II stratum sizes of the designs under these settings. The results were similar with either 25\% or 50\% prevalence of $Z = 1$. The optimal designs favored subjects with $Z=1$, which was partly because ${\rm Var}(X|Z=1)$ was larger than ${\rm Var}(X|Z=0)$, such that  the true value of $X$ was harder to ``guess" when $Z=1$ and validating $X$ among subjects with $Z=1$ was more ``rewarding" than validating $X$ among subjects with $Z=0$.

\begin{table}[!h]
\caption{\label{tab:YXinclZ}Simulation results under outcome and exposure misclassification with an additional error-free covariate.}
\begin{center}
{\tabcolsep=4.25pt
\resizebox{\textwidth}{!}{
\begin{tabular}{@{}rrlrccclrccclrccclrccc@{}}
\hline
\multicolumn{3}{c}{} &  \multicolumn{4}{c}{\textbf{optMLE-2}} & \multicolumn{1}{c}{} & \multicolumn{4}{c}{\textbf{BCC*}} &  \multicolumn{1}{c}{} & \multicolumn{4}{c}{\textbf{CC*}} & \multicolumn{1}{c}{} & \multicolumn{4}{c}{\textbf{SRS}} \\
\cmidrule(l{3pt}r{3pt}){4-7} \cmidrule(l{3pt}r{3pt}){9-12} \cmidrule(l{3pt}r{3pt}){14-17} \cmidrule(l{3pt}r{3pt}){19-22}
$\pmb{\lambda}$ & $\pmb{\beta_Z}$ && \textbf{\% Bias} & \textbf{SE} & \textbf{RE} & \textbf{RI} && \textbf{\% Bias} & \textbf{SE} & \textbf{RE} & \textbf{RI} && \textbf{\% Bias} & \textbf{SE} & \textbf{RE} & \textbf{RI} & &\textbf{\% Bias}& \textbf{SE} & \textbf{RE} & \textbf{RI}\\
\hline
\addlinespace[0.5em]
\multicolumn{22}{c}{$P(Z = 1) = 0.25$} \\
\addlinespace[0.5em]
$-1$ & $-0.25$ &  & $-4.095$ & $0.225$ & $1.329$ & $1.246$ &  & $-1.728$ & $0.296$ & $0.767$ & $0.835$ &  & $-5.322$ & $0.321$ & $0.653$ & $0.833$ &  & $-12.342$ & $0.352$ & $0.543$ & $0.770$\\
& $ 0.00$ &  & $ 0.009$ & $0.227$ & $1.097$ & $1.024$ &  & $-0.734$ & $0.288$ & $0.683$ & $0.792$ &  & $ 3.217$ & $0.325$ & $0.535$ & $0.734$ &  & $ -3.221$ & $0.332$ & $0.515$ & $0.755$\\
& $ 0.25$ &  & $ 0.993$ & $0.221$ & $1.036$ & $0.991$ &  & $ 7.537$ & $0.287$ & $0.613$ & $0.764$ &  & $-1.844$ & $0.333$ & $0.454$ & $0.681$ &  & $  0.261$ & $0.337$ & $0.444$ & $0.682$\\
\addlinespace
$ 0$ & $-0.25$ &  & $-2.982$ & $0.249$ & $0.975$ & $0.960$ &  & $-1.728$ & $0.296$ & $0.690$ & $0.790$ &  & $-5.322$ & $0.321$ & $0.587$ & $0.788$ &  & $-12.342$ & $0.352$ & $0.489$ & $0.729$\\
& $ 0.00$ &  & $-1.351$ & $0.245$ & $1.021$ & $0.965$ &  & $-0.734$ & $0.288$ & $0.735$ & $0.816$ &  & $ 3.218$ & $0.325$ & $0.577$ & $0.756$ &  & $ -3.223$ & $0.332$ & $0.554$ & $0.778$\\
& $ 0.25$ &  & $ 0.023$ & $0.242$ & $0.879$ & $0.925$ &  & $ 7.537$ & $0.287$ & $0.625$ & $0.787$ &  & $-1.844$ & $0.333$ & $0.463$ & $0.701$ &  & $  0.261$ & $0.337$ & $0.453$ & $0.702$\\
\addlinespace
$ 1$ & $-0.25$ &  & $-4.095$ & $0.267$ & $0.943$ & $1.000$ &  & $-1.728$ & $0.296$ & $0.767$ & $0.835$ &  & $-5.322$ & $0.321$ & $0.653$ & $0.833$ &  & $-12.341$ & $0.352$ & $0.543$ & $0.770$\\
& $ 0.00$ &  & $-5.531$ & $0.275$ & $0.840$ & $0.940$ &  & $-0.734$ & $0.288$ & $0.766$ & $0.857$ &  & $ 3.216$ & $0.325$ & $0.601$ & $0.794$ &  & $ -3.221$ & $0.332$ & $0.578$ & $0.816$\\
& $ 0.25$ &  & $-2.354$ & $0.270$ & $0.932$ & $1.065$ &  & $ 7.538$ & $0.287$ & $0.823$ & $0.966$ &  & $-1.844$ & $0.333$ & $0.610$ & $0.860$ &  & $  0.261$ & $0.337$ & $0.596$ & $0.861$\\
\addlinespace[0.5em]
\multicolumn{22}{c}{$P(Z = 1) = 0.50$} \\
\addlinespace[0.5em]
$-1$ & $-0.25$ &  & $-4.512$ & $0.229$ & $1.295$ & $1.241$ &  & $ 0.669$ & $0.304$ & $0.735$ & $0.971$ &  & $-2.470$ & $0.310$ & $0.707$ & $0.913$ &  & $ -3.921$ & $0.338$ & $0.596$ & $0.832$\\
& $ 0.00$ &  & $-3.361$ & $0.228$ & $1.088$ & $1.093$ &  & $-1.451$ & $0.264$ & $0.811$ & $0.902$ &  & $-0.474$ & $0.312$ & $0.579$ & $0.764$ &  & $ -6.093$ & $0.317$ & $0.559$ & $0.764$\\
& $ 0.25$ &  & $-1.931$ & $0.225$ & $0.994$ & $0.972$ &  & $ 4.339$ & $0.287$ & $0.612$ & $0.776$ &  & $-1.055$ & $0.308$ & $0.531$ & $0.685$ &  & $  1.577$ & $0.304$ & $0.545$ & $0.715$\\
\addlinespace
$ 0$ & $-0.25$ &  & $ 0.035$ & $0.250$ & $0.927$ & $1.052$ &  & $ 0.669$ & $0.304$ & $0.628$ & $0.869$ &  & $-2.470$ & $0.310$ & $0.603$ & $0.817$ &  & $ -3.921$ & $0.338$ & $0.509$ & $0.745$\\
& $ 0.00$ &  & $ 4.706$ & $0.239$ & $0.911$ & $0.947$ &  & $-1.451$ & $0.264$ & $0.752$ & $0.838$ &  & $-0.473$ & $0.312$ & $0.538$ & $0.710$ &  & $ -6.092$ & $0.317$ & $0.519$ & $0.710$\\
& $ 0.25$ &  & $ 3.039$ & $0.251$ & $0.833$ & $1.008$ &  & $ 4.339$ & $0.287$ & $0.636$ & $0.838$ &  & $-1.055$ & $0.308$ & $0.553$ & $0.739$ &  & $  1.577$ & $0.304$ & $0.567$ & $0.771$\\
\addlinespace
$ 1$ & $-0.25$ &  & $-4.512$ & $0.277$ & $0.890$ & $1.029$ &  & $ 0.669$ & $0.304$ & $0.735$ & $0.971$ &  & $-2.470$ & $0.310$ & $0.707$ & $0.913$ &  & $ -3.921$ & $0.338$ & $0.596$ & $0.832$\\
& $ 0.00$ &  & $ 0.756$ & $0.267$ & $0.991$ & $0.994$ &  & $-1.452$ & $0.264$ & $1.014$ & $1.012$ &  & $-0.474$ & $0.312$ & $0.725$ & $0.858$ &  & $ -6.093$ & $0.317$ & $0.700$ & $0.857$\\
& $ 0.25$ &  & $ 2.724$ & $0.261$ & $0.916$ & $0.939$ &  & $ 4.339$ & $0.287$ & $0.760$ & $0.867$ &  & $-1.054$ & $0.308$ & $0.660$ & $0.765$ &  & $  1.577$ & $0.304$ & $0.678$ & $0.799$\\
\hline
\end{tabular}}}
\end{center}
{\noindent \% Bias and SE are, respectively, the empirical percent bias and standard error of the MLE. The grid search algorithm successfully located the optMLE and optMLE-2 designs in all and $>99\%$ of replicates per setting, respectively; 16 ($<0.1\%$) problematic replicates out of \num{18000} were discarded. All other entries are based on \num{1000} replicates.
\par}
\end{table}

\clearpage

\begin{table}[bt]
\caption{\label{tab:mod_misspec}Additional simulation results for other designs when the misspecification models used in the optimal design can be misspecified.}
\begin{center}
\resizebox{\linewidth}{!}{
{\tabcolsep=1.9pt
\begin{tabular}{rrlrccclrccclrccc}
\hline
\multicolumn{3}{c}{} & 
\multicolumn{4}{c}{\textbf{BCC*}} &  \multicolumn{1}{c}{} & \multicolumn{4}{c}{\textbf{CC*}} & \multicolumn{1}{c}{} & \multicolumn{4}{c}{\textbf{SRS}} \\
\cmidrule(l{3pt}r{3pt}){4-7}\cmidrule(l{3pt}r{3pt}){9-12}\cmidrule(l{3pt}r{3pt}){14-17}
$\pmb{\delta_1}$ & $\pmb{\delta_2}$ && 
\textbf{\% Bias} & \textbf{SE} & \textbf{RE} & \textbf{RI} && 
\textbf{\% Bias} & \textbf{SE} & \textbf{RE} & \textbf{RI} && 
\textbf{\% Bias} & \textbf{SE} & \textbf{RE} & \textbf{RI} \\
\hline
\addlinespace[0.5em]
\multicolumn{17}{c}{\textbf{Misspecified misclassification mechanism for $\pmb{Y^*}$ and $\pmb{X^*}$}} \\
\addlinespace[0.5em]
$-1.0$ & $-1.0$ &  & $ -0.570$ & $0.317$ & $0.754$ & $0.893$ &  & $ 2.058$ & $0.336$ & $0.672$ & $0.804$ &  & $ 1.369$ & $0.364$ & $0.572$ & $0.783$\\
$-0.5$ & $-0.5$ &  & $ -0.053$ & $0.310$ & $0.674$ & $0.903$ &  & $ 2.121$ & $0.326$ & $0.610$ & $0.819$ &  & $-7.261$ & $0.348$ & $0.537$ & $0.759$\\
$ 0.0$ & $ 0.0$ &  & $  7.538$ & $0.287$ & $0.823$ & $0.966$ &  & $-1.844$ & $0.333$ & $0.610$ & $0.860$ &  & $ 0.261$ & $0.337$ & $0.596$ & $0.861$\\
$ 0.5$ & $ 0.5$ &  & $  5.243$ & $0.318$ & $0.657$ & $0.790$ &  & $-1.785$ & $0.318$ & $0.656$ & $0.805$ &  & $-2.903$ & $0.358$ & $0.519$ & $0.710$\\
$ 1.0$ & $ 1.0$ &  & $-10.917$ & $0.320$ & $0.599$ & $0.779$ &  & $-1.241$ & $0.339$ & $0.533$ & $0.712$ &  & $-7.684$ & $0.339$ & $0.533$ & $0.710$\\
\addlinespace[0.5em]
\multicolumn{17}{c}{\textbf{Misspecified misclassification mechanism for $\pmb{Y^*}$}} \\
\addlinespace[0.5em]
$0.0$ & $-1.0$ &  & $ 3.504$ & $0.314$ & $0.712$ & $0.795$ &  & $-3.587$ & $0.336$ & $0.622$ & $0.749$ &  & $-6.257$ & $0.344$ & $0.595$ & $0.812$\\
$0.0$& $-0.5$ &  & $-1.159$ & $0.315$ & $0.806$ & $0.912$ &  & $ 2.174$ & $0.346$ & $0.671$ & $0.775$ &  & $ 1.268$ & $0.363$ & $0.610$ & $0.782$\\
$0.0$& $ 0.0$ &  & $ 7.538$ & $0.287$ & $0.823$ & $0.966$ &  & $-1.844$ & $0.333$ & $0.610$ & $0.860$ &  & $ 0.261$ & $0.337$ & $0.596$ & $0.861$\\
$0.0$& $ 0.5$ &  & $ 0.570$ & $0.312$ & $1.011$ & $1.001$ &  & $-3.129$ & $0.323$ & $0.946$ & $0.993$ &  & $-2.122$ & $0.347$ & $0.817$ & $0.890$\\
$0.0$& $ 1.0$ &  & $-6.256$ & $0.306$ & $0.716$ & $0.884$ &  & $ 2.449$ & $0.336$ & $0.597$ & $0.733$ &  & $-8.055$ & $0.354$ & $0.536$ & $0.719$\\
\addlinespace[0.5em]
\multicolumn{17}{c}{\textbf{Misspecified misclassification mechanism for $\pmb{X^*}$}} \\
\addlinespace[0.5em]
$-1.0$ & $0.0$ &  & $-11.340$ & $0.299$ & $0.837$ & $0.968$ &  & $-7.776$ & $0.333$ & $0.674$ & $0.804$ &  & $-2.565$ & $0.333$ & $0.675$ & $0.846$\\
$-0.5$ & $0.0$ &  & $ -2.549$ & $0.318$ & $0.625$ & $0.807$ &  & $-7.388$ & $0.330$ & $0.583$ & $0.750$ &  & $-9.968$ & $0.343$ & $0.538$ & $0.752$\\
$ 0.0$ & $0.0$ &  & $  7.538$ & $0.287$ & $0.823$ & $0.966$ &  & $-1.844$ & $0.333$ & $0.610$ & $0.860$ &  & $ 0.261$ & $0.337$ & $0.596$ & $0.861$\\
$ 0.5$ & $0.0$ &  & $ -2.139$ & $0.316$ & $0.646$ & $0.836$ &  & $ 0.820$ & $0.340$ & $0.559$ & $0.783$ &  & $-3.184$ & $0.344$ & $0.545$ & $0.786$\\
$ 1.0$ & $0.0$ &  & $ -2.764$ & $0.326$ & $0.638$ & $0.784$ &  & $ 6.405$ & $0.322$ & $0.653$ & $0.759$ &  & $-1.454$ & $0.342$ & $0.581$ & $0.803$\\
\hline
\end{tabular}}}    
\end{center}
{\noindent The error-prone exposure and outcome were generated from $P(X^*=1|Y,X,Z) = [1 + \exp\{-(-1.1 + 0.45 Y + 2.2 X + Z + \delta_1 X Z)\}]^{-1}$ and $P(Y^*=1|X^*,Y,X,Z) = [1 + \exp\{-(-1.1 + 0.275X^* + 2.2 Y + 0.275 X + Z + \delta_{2} X Z)\}]^{-1}$, respectively. \% Bias and SE are, respectively, the empirical percent bias and standard error of the MLE. All entries are based on \num{1000} replicates.
\par}
\end{table}

\begin{figure}[h]
    \centering
	   \centering
	   \includegraphics[width=\textwidth]{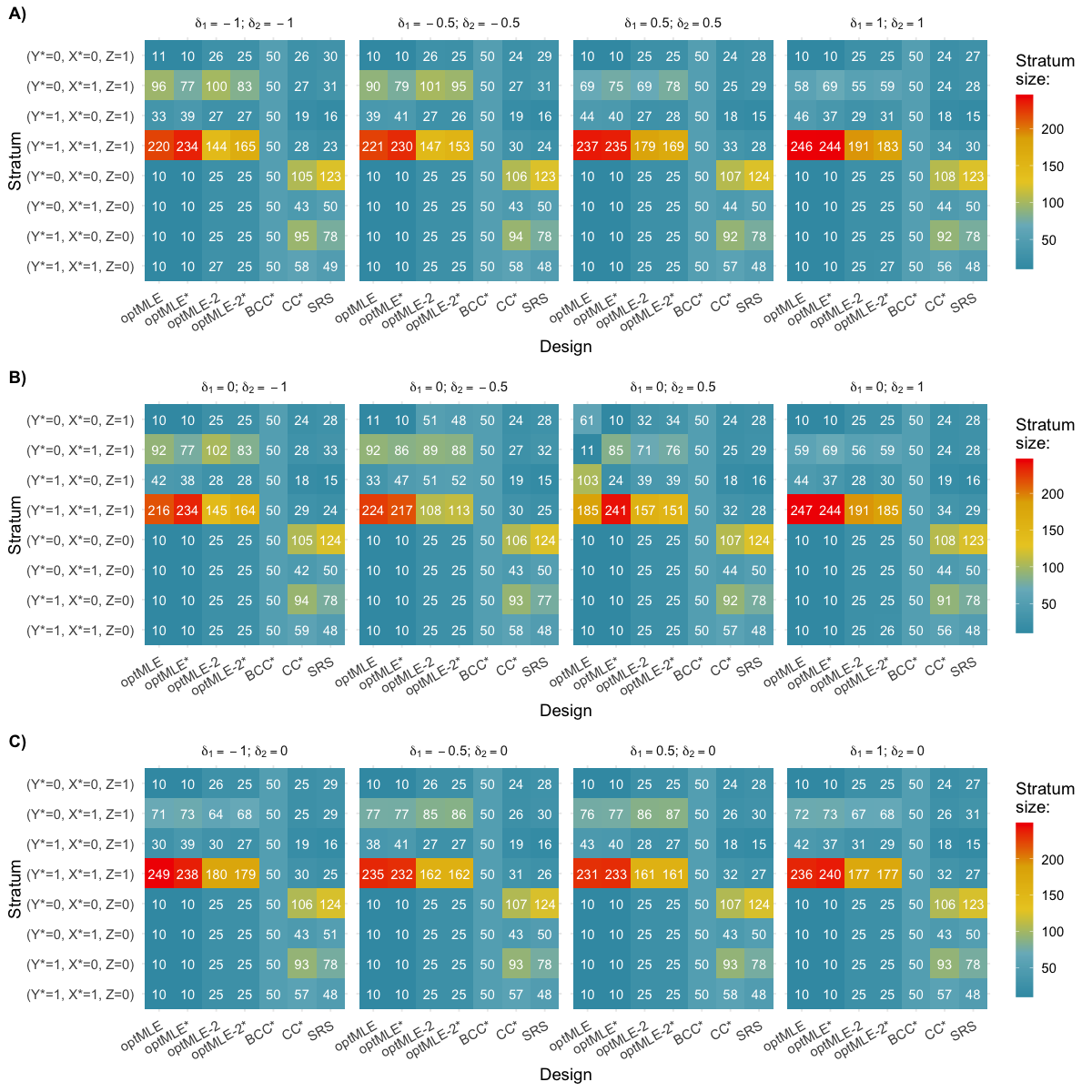}
    \caption{Average Phase II stratum sizes $n_{y^*x^*z}$ when optimal designs optMLE* and optMLE-2* can be derived based on misspecified misclassification mechanisms. In \textbf{A)}, \textbf{B)}, and \textbf{C)}, the $Y^*$ and $X^*$, $Y^*$ only, and $X^*$ only misclassification mechanisms can be misspecified, respectively.}
    \label{fig:des_misspec}
\end{figure}

\subsection{Outcome Misclassification Only}\label{AppC_Yonly}

For the special scenario of outcome misclassification alone, $X^* = X$ such that Equation (2.1) reduces to $P(Y^*,Y,X) = P(Y^*|Y,X)P(Y|X)P(X)$. We generated $Y$ and $X$ in the same way as in Section~3.2, with {$p_{y0} = 0.3$} and $p_{x} = 0.1$, for a sample of $N = \num{10000}$ subjects. We generated error-prone $Y^*$ from a Bernoulli distribution with $P(Y^*=1|Y,X) = [1 + \exp\{-(\alpha_0 + \alpha_1 Y + 0.28 X)\}]^{-1}$, where $\alpha_0$ and $\alpha_1$ were defined in the same way as in Section~3.2 with $FPR_{0}(Y^*)\in \{0.1, 0.5\}$ and $TPR_{0}(Y^*)\in \{0.9, 0.5\}$. {Note that without $X^*$, the baseline false positive and true positive rates for $Y^*$ are defined with $X=0$ in this setting (hence the single zero subscript).} We set $n = 400$. Without exposure misclassification, the sampling strata for the BCC*, optMLE, and optMLE-2 designs were defined by $(Y^*, X)$. Each setting was replicated \num{1000} times.

\begin{figure}
    \centering
	   \centering
	   \includegraphics[width=\textwidth]{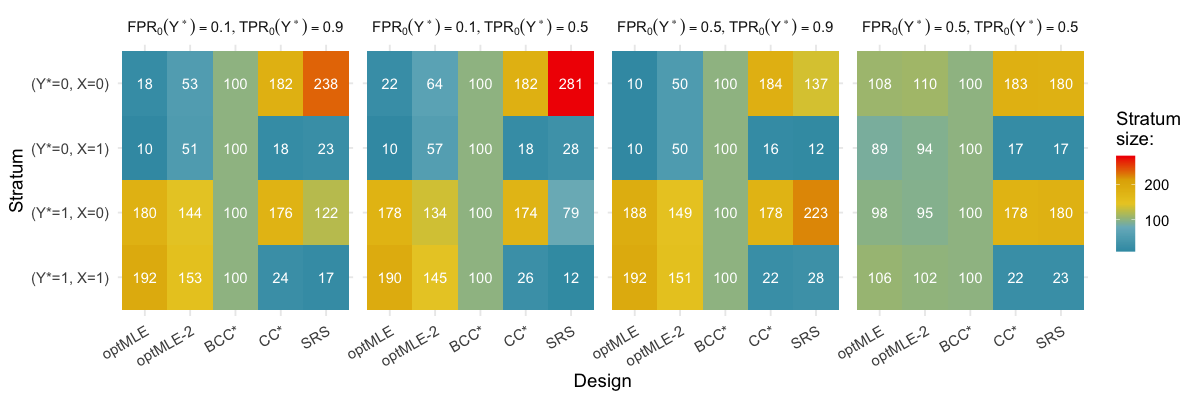}
    \caption{Average Phase II stratum sizes $n_{y^*x}$ under outcome misclassification.}
    \label{fig:errors_Y}
\end{figure}

Simulation results for the MLE are included in Table~\ref{tab:errors_Yonly_Xonly}(a). The optMLE-2 design did not lose much efficiency to the optMLE design and typically surpassed the efficiencies of the BCC*, CC*, and SRS designs, with gains as high as 21\%, 72\%, and 74\%, respectively. Fig.~\ref{fig:errors_Y} shows the average Phase II stratum sizes selected under each of the designs. The optimal designs favored strata with the less-frequent value of $Y^*$ (i.e., $Y^* = 1$) in all settings where it was informative (i.e., $FPR_{0}(Y^*) \neq 0.5$ or $TPR_{0}(Y^*) \neq 0.5$). In the highest error setting, the optimal designs appeared to be similar to the BCC* design. 

\begin{table}[!h]
\caption{\label{tab:errors_Yonly_Xonly}Simulation results under outcome or exposure misclassification only.}
\begin{center}
{\tabcolsep=4.25pt
\resizebox{\textwidth}{!}{
\begin{tabular}{@{}cclrccclrccclrccclrccc@{}}
\hline\\
\addlinespace[0.5em]
\multicolumn{22}{c}{\textbf{(a) Outcome Misclassification Only}} \\
\addlinespace[0.5em]
\multicolumn{2}{c}{\textbf{Misclassification}} & \multicolumn{1}{c}{} &  \multicolumn{4}{c}{\textbf{optMLE-2}} & \multicolumn{1}{c}{} & \multicolumn{4}{c}{\textbf{BCC*}} &  \multicolumn{1}{c}{} & \multicolumn{4}{c}{\textbf{CC*}} & \multicolumn{1}{c}{} & \multicolumn{4}{c}{\textbf{SRS}} \\
\cmidrule(l{3pt}r{3pt}){1-2}\cmidrule(l{3pt}r{3pt}){4-7} \cmidrule(l{3pt}r{3pt}){9-12} \cmidrule(l{3pt}r{3pt}){14-17} \cmidrule(l{3pt}r{3pt}){19-22}
\textbf{$\pmb{\FPR_{0}(Y^*)}$} & \textbf{$\pmb{\TPR_{0}(Y^*)}$} && \textbf{\% Bias} & \textbf{SE} & \textbf{RE} & \textbf{RI} && \textbf{\% Bias} & \textbf{SE} & \textbf{RE} & \textbf{RI} && \textbf{\% Bias} & \textbf{SE} & \textbf{RE} & \textbf{RI} && \textbf{\% Bias} & \textbf{SE} & \textbf{RE} & \textbf{RI} \\
\hline
$0.1$ & $0.9$ &  & $ 2.066$ & $0.119$ & $0.935$ & $0.965$ &  & $-0.752$ & $0.133$ & $0.741$ & $0.800$ &  & $ 1.635$ & $0.209$ & $0.302$ & $0.526$ &  & $0.881$ & $0.234$ & $0.240$ & $0.452$\\
& $0.5$ &  & $ 1.158$ & $0.179$ & $0.866$ & $0.991$ &  & $-0.696$ & $0.192$ & $0.754$ & $0.902$ &  & $-2.929$ & $0.293$ & $0.325$ & $0.563$ &  & $3.086$ & $0.315$ & $0.281$ & $0.548$\\
$0.5$ & $0.9$ &  & $-0.081$ & $0.199$ & $0.859$ & $0.898$ &  & $-0.494$ & $0.218$ & $0.717$ & $0.790$ &  & $-4.366$ & $0.376$ & $0.241$ & $0.477$ &  & $2.088$ & $0.345$ & $0.287$ & $0.519$\\
& $0.5$ &  & $ 3.171$ & $0.219$ & $0.914$ & $0.997$ &  & $ 1.461$ & $0.207$ & $1.019$ & $1.061$ &  & $-2.941$ & $0.373$ & $0.314$ & $0.623$ &  & $0.319$ & $0.348$ & $0.362$ & $0.657$\\
\addlinespace[0.5em]
\multicolumn{22}{c}{\textbf{(b) Exposure Misclassification Only}} \\
\addlinespace[0.5em]
\multicolumn{2}{c}{\textbf{Misclassification}} & \multicolumn{1}{c}{} &  \multicolumn{4}{c}{\textbf{\textbf{optMLE-2}}} & \multicolumn{1}{c}{} & \multicolumn{4}{c}{\textbf{\textbf{BCC*}}} &  \multicolumn{1}{c}{} & \multicolumn{4}{c}{\textbf{\textbf{CC*}}} & \multicolumn{1}{c}{} & \multicolumn{4}{c}{\textbf{\textbf{SRS}}} \\
\cmidrule(l{3pt}r{3pt}){1-2}\cmidrule(l{3pt}r{3pt}){4-7} \cmidrule(l{3pt}r{3pt}){9-12} \cmidrule(l{3pt}r{3pt}){14-17} \cmidrule(l{3pt}r{3pt}){19-22}
\textbf{$\pmb{\FPR_{0}(X^*)}$} & \textbf{$\pmb{\TPR_{0}(X^*)}$}  && \textbf{\% Bias} & \textbf{SE} & \textbf{RE} & \textbf{RI} && \textbf{\% Bias} & \textbf{SE} & \textbf{RE} & \textbf{RI} && \textbf{\% Bias} & \textbf{SE} & \textbf{RE} & \textbf{RI} && \textbf{\% Bias} & \textbf{SE} & \textbf{RE} & \textbf{RI} \\
\hline
$0.1$ & $0.9$ &  & $-1.716$ & $0.158$ & $0.861$ & $0.917$ &  & $3.146$ & $0.180$ & $0.660$ & $0.838$ &  & $ 1.821$ & $0.282$ & $0.270$ & $0.536$ &  & $ 1.134$ & $0.285$ & $0.264$ & $0.540$\\
& $0.5$ &  & $ 0.892$ & $0.206$ & $1.002$ & $0.928$ &  & $4.337$ & $0.237$ & $0.754$ & $0.788$ &  & $ 2.750$ & $0.313$ & $0.433$ & $0.607$ &  & $ 0.785$ & $0.332$ & $0.385$ & $0.618$\\
$0.5$ & $0.9$ &  & $-3.594$ & $0.298$ & $0.852$ & $0.902$ &  & $2.773$ & $0.361$ & $0.580$ & $0.736$ &  & $-3.331$ & $0.327$ & $0.706$ & $0.831$ &  & $-4.604$ & $0.336$ & $0.667$ & $0.795$\\
& $0.5$ &  & $-0.987$ & $0.345$ & $0.941$ & $0.957$ &  & $5.740$ & $0.343$ & $0.949$ & $0.959$ &  & $ 2.027$ & $0.348$ & $0.923$ & $0.964$ &  & $-3.840$ & $0.362$ & $0.854$ & $0.880$\\
\hline
\end{tabular}}}    
\end{center}
{\noindent \% Bias and SE are, respectively, the empirical percent bias and standard error of the MLE. The grid search algorithm successfully located the optMLE and optMLE-2 designs in all settings and replicates. Each entry is based on 1000 replicates.
\par}
\end{table}

\subsection{Exposure Misclassification Only}\label{AppC_Xonly}

For the special scenario of exposure misclassification alone, $Y^* = Y$ such that Equation (2.1) reduces to $P(X^*,Y,X) = P(X^*|Y,X)P(Y|X)P(X)$. We generated $Y$ and $X$ in the same way as in \ref{AppC_Yonly} for a Phase I sample of $N = \num{10000}$ subjects. We generated error-prone $X^*$ from a Bernoulli distribution with $P(X^* = 1|Y,X) = [1 + \exp\{-(\gamma_0 + 0.45 Y + \gamma_1 X)\}]^{-1}$, where $\gamma_0$ and $\gamma_1$ were defined in the same way as in Section~3.2 with $\FPR_{0}(X^*) \in \{0.1, 0.5\}$ and $\TPR_{0}(X^*) \in \{0.9, 0.5\}$. We set $n = 400$. Without outcome misclassification, the sampling strata for the BCC*, optMLE, and optMLE-2 designs were defined by $(Y, X^*)$. Results in each setting are based on \num{1000} replicates. 

\begin{figure}[!h]
    \centering
	   \centering
	   \includegraphics[width=\textwidth]{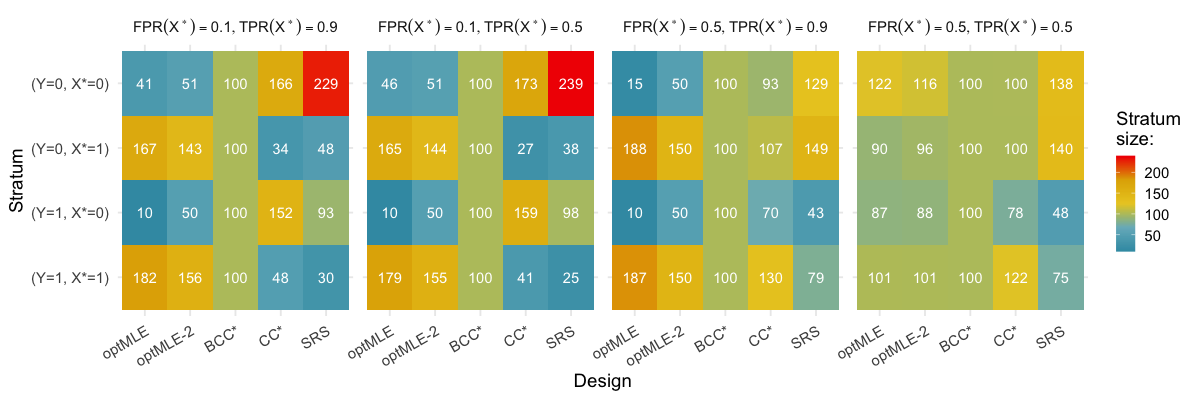}
    \caption{Average Phase II stratum sizes $n_{yx^*}$ under exposure misclassification.}
    \label{fig:errors_X}
\end{figure}

Simulation results for the MLE are included in Table~\ref{tab:errors_Yonly_Xonly}(b). The optMLE-2 design did not lose much efficiency to the optMLE design and typically surpassed the efficiencies of the BCC*, CC*, and SRS designs, with gains as high as 32\%, 69\% and 69\%, respectively. Fig.~\ref{fig:errors_X} shows the average Phase II stratum sizes selected under each of the designs. The optimal designs favored strata with the less-frequent value of $X^*$ (i.e., $X^* = 1$) in all settings where it was informative (i.e., $\FPR_{0}(X^*) \neq 0.5$ or $\TPR_{0}(X^*) \neq 0.5$). In the highest error setting, the optimal designs appeared to be similar to the BCC* design. Together with \ref{AppC_Yonly}, these results suggest that the optimal designs seemed to target the less-frequent value of the error-prone variable with very little regard for the error-free variable. 

\clearpage 

\begin{table}[!p]
\caption{\label{tab:hist_audit_ccasanet}Historic TB audit results in CCASAnet. Note: No subject had both outcome and exposure misclassification.}
{\tabcolsep=4.25pt
\begin{tabular}{@{}rlcccc@{}}
\hline
\multicolumn{4}{c}{} & \multicolumn{2}{c}{\textbf{Misclassified (\%)}} \\
\cmidrule(l{3pt}r{3pt}){5-6}
\textbf{Country} & & \textbf{Audited} & & \textbf{Treatment Completion ($Y^*$)} & \textbf{Bacterial Confirmation ($X^*$)} \\
\hline
\addlinespace[0.5em]
\multicolumn{6}{l}{Country Grouping = 0}\\
\addlinespace[0.5em]
A && 6 && 2 ($33.3$\%) & 0 ($0.0$\%)\\
B && 7 && 0 ($0.0$\%) & 2 ($28.6$\%) \\
\addlinespace[0.5em]
\multicolumn{6}{l}{Country Grouping = 1}\\
\addlinespace[0.5em]
C && 6 &&  1 ($16.7$\%) & 1 ($16.7$\%)\\
D && 10 && 2 ($20.0$\%) & 5 ($50.0$\%) \\
\addlinespace[0.5em]
\multicolumn{6}{l}{Country Grouping = 2}\\
\addlinespace[0.5em]
E && 4 && 0 ($0.0$\%) & 0 ($0.0$\%) \\
\hline 
\end{tabular}
}
\end{table}

\begin{table}[!p]
\caption{\label{tab:hist_models_ccasanet}Parameter estimates for TB analysis in CCASAnet using historic audits.}
\begin{center}
{\tabcolsep=4.25pt
\begin{tabular}[t]{@{}llr@{}}
\hline
& \textbf{Coefficient} & \textbf{log OR} \\
\hline
\addlinespace[0.5em]
\multicolumn{3}{l}{\textbf{Analysis model ($Y$)}}\\
\addlinespace[0.5em]
& Intercept & $0.752$ \\
& $X$ & $-0.415$ \\
& CoG = 0 (Co = A--B) & Referent \\
& CoG = 1 (Co = C--D) & $0.601$ \\
& CoG = 2 (Co = E) & $0.211$ \\
\addlinespace[0.5em]
\multicolumn{3}{l}{\textbf{Outcome misclassification mechanism ($Y^*$)}}\\
& Intercept & $2.088$ \\
& $X^*$ & $0.156$ \\
& $Y^*$ & $4.644$ \\
& $X$ & $2.485$ \\
& CoG = 0 (Co = A--B) & Referent \\
& CoG = 1 (Co = C--D) & $-1.182$ \\
& CoG = 2 (Co = E) & $-0.956$ \\
\addlinespace[0.5em]
\multicolumn{3}{l}{\textbf{Exposure misclassification mechanism ($X^*$)}}\\
& Intercept & $-0.600$ \\
& $Y$ & $-2.611$ \\
& $X$ & $4.770$ \\
& CoG = 0 (Co = A--B) & Referent \\
& CoG = 1 (Co = C--D) & $1.685$ \\
& CoG = 2 (Co = E) & $0.170$ \\
\addlinespace[0.5em]
\multicolumn{3}{l}{\textbf{Exposure model ($X$)}}\\
& Intercept & $-1.017$ \\
& CoG = 0 (Co = A--B) & Referent \\
& CoG = 1 (Co = C--D) & $-0.160$ \\
& CoG = 2 (Co = E) & $-0.592$ \\
\hline
\end{tabular}}    
\end{center}
{\noindent The optimal designs for CCASAnet were based on historic parameters $\hat{\beta}^{(h)} = -0.415$ and $\hat{\pmb{\eta}}^{(h)\rm T}$ = ($0.752$, $0$, $0.601$, $0.601$, $0.211$, $2.088$, $0.156$, $4.644$, $2.485$, $0$, $-1.182$, $-1.182$, $-0.956$, $-0.6$, $-2.611$, $4.77$, $0$, $1.685$, $1.685$, $0.17$, $-1.017$, 0, $-0.16$, $-0.16$, $-0.592)^{\rm T}$.\par}
\end{table}

\begin{table}[!p]
\caption{\label{tab:other_estimators}Simulation results comparing the MLE and SMLE.}
\begin{center}
{\tabcolsep=4.25pt
\begin{tabular}{@{}rlrccclrccc@{}}
\hline
\multicolumn{2}{c}{\textbf{ }} & \multicolumn{4}{c}{\textbf{MLE}} & \multicolumn{1}{c}{\textbf{ }} & \multicolumn{4}{c}{\textbf{SMLE}} \\
\cmidrule(l{3pt}r{3pt}){3-6} \cmidrule(l{3pt}r{3pt}){8-11}
\textbf{Design} & \textbf{ } & \textbf{\% Bias} & \textbf{SE} & \textbf{RE} & \textbf{RI} & \textbf{ } & \textbf{\% Bias} & \textbf{SE} & \textbf{RE} & \textbf{RI} \\
\hline
optMLE && $3.333$ & $0.176$ & $1.000$ & $1.000$ && $2.667$ & $0.176$ & $1.000$ & $1.000$ \\
optMLE-2 && $1.000$ & $0.181$ & $0.942$ & $0.933$ && $0.333$ & $0.181$ & $0.946$ & $0.934$ \\
BCC* && $0.667$ & $0.198$ & $0.788$ & $0.853$ && $0.333$ & $0.198$ & $0.785$ & $0.855$ \\
CC* && $0.000$ & $0.270$ & $0.426$ & $0.616$ && $0.000$ & $0.270$ & $0.425$ & $0.618$ \\
SRS && $-5.667$ & $0.305$ & $0.333$ & $0.541$ && $-5.667$ & $0.305$ & $0.332$ & $0.544$ \\
\hline
\end{tabular}}    
\end{center}
{\noindent The SMLE was proposed with $X^*$ as a surrogate for $X$ such that $(Y \perp X^*)|X$. Thus, $X^*$ was generated from a Bernoulli distribution with $P(X^*=1|Y,X) = [1 + \exp\{-(\gamma_0 + \gamma_1 X )\}]^{-1}$. All other variables were generated as in Section~3.2, with $p_{y0} = 0.3$, $p_{x} = 0.1$, $\FPR_{00}(Y^*) = 0.1$, $\TPR_{00}(Y^*) = 0.9$, $\FPR(X^*) = 0.1$, and $\TPR(X^*) = 0.9$. \% Bias and SE are, respectively, the empirical percent bias and standard error of the estimators. Each entry is based on \num{1000} replicates.\par}
\end{table}